\documentclass{an}
\usepackage{graphicx}
\usepackage{times}
\Volume{01}              
\Year{2005}              
\Month{08}               
\Pagespan{1}{7}      	 

\begin{document}
\lhead[\thepage]{A.N. Bilir et al.: Galactic model parameters for field giants}
\rhead[Astron. Nachr./AN~{\bf XXX} (200X) X]{\thepage}
\headnote{Astron. Nachr./AN {\bf 32X} (200X) X, XXX--XXX}

\title{Galactic model parameters for field giants separated 
from field dwarfs by their 2MASS and V apparent magnitudes}

\author{S. Bilir
\and  S. Karaali
\and  T. G\"uver
\and  Y. Karata\c{s} 
\and  S. G. Ak}
\institute{Istanbul University Science Faculty, 
           Department of Astronomy and Space Sciences, 
           34119, University-Istanbul, Turkey}
\date{} 

\abstract{We present a procedure which separates field dwarfs and 
field giants by their 2MASS and V apparent magnitudes. The procedure is 
based on the spectroscopically selected standards, hence it is confident. We 
applied this procedure to  stars in two fields, SA 54 and SA 82, and we 
estimated a full set of Galactic model parameters for giants including their 
total local space density. Our results are in agreement  with the ones appeared 
recently in the literature.     
}

\correspondence{sbilir@istanbul.edu.tr}

\maketitle

\section{Introduction}

For some years, there has been a conflict among the researchers about the 
history of our Galaxy. Yet there has been a large improvement about this 
topic since the pioneering work of Eggen, Lynden-Bell \& Sandage (1962) 
who argued that the Galaxy collapsed in a free-fall time ($\sim2\times10^{8}$ 
yr). Now, we know that the Galaxy collapsed over many Gyr (e.g. Yoshii \& 
Saio 1979; Norris, Bessell \& Pickles 1985; Norris 1986; Sandage \& Fouts 
1987; Carney, Latham \& Laird 1990; Norris \& Ryan 1991; Beers \& Sommer-Larsen 
1995) and at least some of its components are formed from the merger or 
accretion of numerous fragments, such as dwarf-type galaxies (cf. Searle \& 
Zinn 1978; Freeman \& Band-Hawthorn 2002, and references therein). Also the 
number of population components increased from two to three, complicating 
interpretations of any data set. The new component (the thick disc) was 
introduced by Gilmore \& Reid (1983) in order to explain the observation 
that star counts towards the South Galactic Pole were not in agreement with 
a single-disc (thin disc) component, but rather could be much better 
represented by two such components. The new component is discussed by 
Gilmore \& Wyse (1985) and Wyse \& Gilmore (1986).

The researchers use different methods to determine the parameters for three 
population components and try to interpret them in relation to the formation 
and evolution of the Galaxy. Among the parameters, the local density and the 
scaleheight of the thick disc are the ones for which the numerical values 
improved relative to the original ones claimed by Gilmore \& Reid (1983). In 
fact, the researchers indicate a tendency for the original local density of the 
thick disc to increase from 2 to 10 per cent relative to the local density and 
for its scaleheight to decrease from the original value of 1.45 kpc down to 0.65 
kpc (Chen et al. 2001). In some studies, the range of values for the parameters 
is large, especially for the thick disc. For example, Chen et al. (2001) and 
Siegel et al. (2002) give 6.5-13 and 6-10 per cent, respectively, for the 
relative local density for the thick disc. We showed that the model parameters 
are absolute magnitude dependent, and that such a process limits the range of 
the parameters considerably (Karaali, Bilir \& Hamzao\u glu, 2004a).

The studies related to the Galactic structure are usually carried out by star 
counts. However, it is stated by many authors (cf. Siegel et al., 2002) that 
the non-invertibility and the vagaries of solving the non-unique convolution 
by trial and error limit the star counts and will be a weak tool for exploring 
the Galaxy. Direct comparison between the theoretical and observed space 
densities is another method used. In the literature, there is limited number 
of research based on this method. The works of Basle group (del Rio \& Fenkart 
1987; Fenkart \& Karaali 1987) and the recent work of Phleps et al. (2000), 
Siegel et al. (2002), Karaali et al. (2003), Du et al. (2003), Karaali et 
al. (2004a) and Bilir, Karaali \& Gilmore (2005a) can be given as examples of 
this research.

\begin{figure*}
\resizebox{17cm}{5.51cm}{\includegraphics*{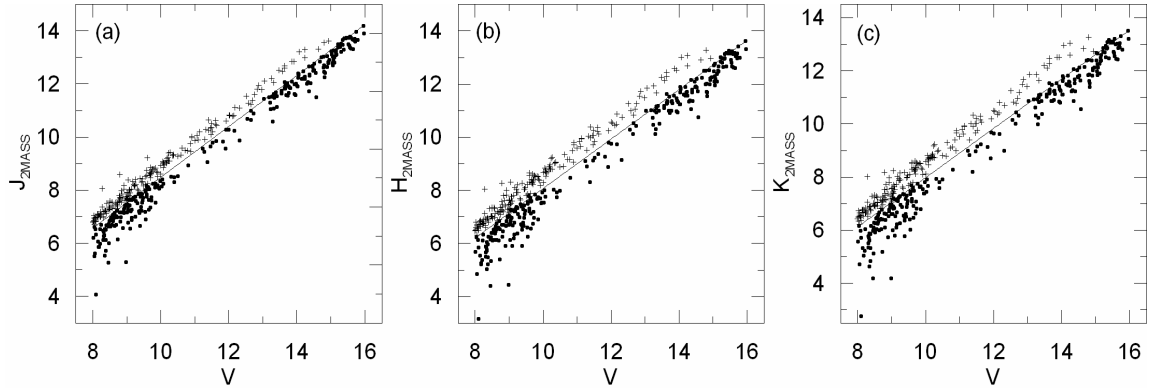}} 
\caption {V and 2MASS magnitudes of the Cayrel et al. (2001) and Ratnatunga 
\& Freeman (1989) in our sample. (a) $J/V$, (b) $H/V$ and (c) $K/V$. The 
symbols ($+$) and ($\bullet$) correspond to dwarfs and giants, respectively.}
\end {figure*}

In many studies, the Galactic model parameters are estimated without any 
discrimination between dwarf and giants, whereas some researchers estimated model 
parameters for different star categories (e.g. Pritchet 1983, Bahcall \& Soneira 
1984, Buser \& Kaeser 1985 and Mendez \& Altena 1996). A very recent work is 
devoted only to estimation of the model parameters for giants (Cabrera-Lavers, 
Garzon \& Hammersley 2005). Separation of field giants and dwarfs plays an 
important role in the Galactic model estimation. The most efficient method for 
separation of stars into these categories is of course spectroscopical one. It 
can be done either by inspection of their spectral lines or using their surface 
gravities. However, both procedures are rather tiring. An easier procedure is 
to separate dwarfs and evolved stars (subgiants or giants) such as to obtain a 
luminosity function consistent with the local luminosity function of nearby 
stars due to Gliese \& Jahreiss (1991) and Jahreiss \& Wielen (1997). The 
procedure of this separation is based on the fact that the local luminosity 
functions obtained for many fields indicated a systematic excess of star counts 
relative to the luminosity function of nearby stars for the fainter segment, 
i.e. $M(V)\geq5^{m}.5$, and a deficit for brighter segment, $M(V)<5^{m}.5$. 
(in RGU system $M(G)\geq6^{m}$ and $M(G)<6^{m}$, respectively). The works of 
Karaali (1992); Ak, Karaali \& Buser (1998); Karata\c{s}, Karaali \& Buser 
(2001); Karaali et al. (2004b); Bilir, Karaali \& Buser (2004); and Karata\c{s} 
et al. (2004) can be given as examples for application of this procedure.  

In this work, we present a different procedure for separation of field dwarfs 
and giants by their Two Micron All Sky Survey (2MASS), i.e. $J$, $H$ and $K$, 
and $V$ magnitudes down to the limiting magnitude of $V=16$. The standard stars 
used for defining the procedure are selected spectroscopically, hence our 
procedure would be applied confidently. The resultant Galactic model parameters 
for giants in two fields, SA 54 and SA 82, where their separation from the dwarfs 
is carried out by this procedure confirm our suggestion. The description of the 
procedure is given in Section 2, Section 3 is devoted to the application of the 
procedure to the giants in SA 54 and SA 82, the Galactic model parameters for 
giants estimated for two fields is given in Section 4, and finally Section 5 
provides a conclusion.              

\section{Procedure for separation of field giants from field dwarfs}

We used 196 field dwarfs and 156 field giants, taken from the catalogs of 
Cayrel, Soubiran and Ralite (2001) to obtain a procedure which separates 
the field dwarfs and giants. The double or multiple and variable stars, 
and stars which were exposed to interstellar reddening 
($|b|\leq 20^{\circ}$) in the Cayrel et al. (2001) catalogue have been 
eliminated. According to these criterias, totally 352 stars  were 
separated into two categories, i.e. dwarf and giant, by us 
according to their surface gravities. Thus, stars with $\log g\leq3$ 
were classified as giants, whereas those with $\log g\geq4$ were assumed to be 
dwarfs. The catalog of Ratnatunga \& Freeman (1989) from the other hand, 
offers 101 giants classified spectroscopically. The apparent limiting 
magnitude for stars in this catalog is $V=16$. Thus, we have a sample of 453 
stars separated into 196 dwarfs and 257 giants confidently.

Our aim is to compare the 2MASS apparent magnitudes, $J$, $H$ and $K$, with 
the $V$ apparent magnitude, from which we expect a systematic deviation 
between the two stellar categories in the two magnitude plane. The 2MASS 
(Skrutskie et al. 1997) is using two 1.3m telescopes, one on Mt. Hopkins in 
Arizona and one at the Cerro Tololo Inter-American Observatory in Chile, to 
survey the entire sky in near-infrared light\footnote{http://www.ipac.caltech.edu/
2mass/overview/about2mass.html}. In addition providing a context for the 
interpretation of results obtained at infrared and other wavelengths, 2MASS 
will provide direct answers to immediate questions on the large-scale 
structure of the Milky Way and the Local Universe. We used the 2MASS All-Sky 
Catalog of Point Sources of Cutri et al. (2003) to draw the $J$, $H$ and $K$ 
magnitudes for 453 stars mentioned above as well as their $V$ magnitudes. The 
data are given in Table 1. Fig. 1a-c compares the 2MASS magnitudes and the $V$ 
magnitude for the sample stars where a systematic deviation between giants and 
dwarfs is conspicuous, especially at the faint magnitudes. We adopted the 
straight line defined by positions of two giants as the upper envelope of 
giants (hence lower envelope of dwarfs). The first position corresponds to the 
giant faintest in two magnitudes, whereas the other one is the giant brightest 
in $V$ but faintest in the 2MASS magnitude. The equation for three lines in 
Fig. 1a-c are given in the following:

\begin{equation}
J = 0.957V - 1.079
\end{equation}

\begin{equation}
H = 0.931V - 1.240
\end{equation}

\begin{equation}
K = 0.927V - 1.292	
\end{equation}

\setcounter{table}{1}
\begin{table}
\center
\caption{Data for the fields SA 54 and SA 82. The symbols give: $N_{s}$: number 
of sources, $N_{glx}$: number of galaxies, $N_{d}$: number of dwarfs and $N_{g}$: 
number of giants.}
\begin{tabular}{cccccccc}
\hline
Field & l        & b       &  Size        & $N_{s}$ & $N_{glx}$ & $N_{d}$ & $N_{g}$\\
      & ($^{o}$) & ($^{o}$)&  (deg$^{2}$) &         &           &         &        \\

\hline
SA 54 & 200.1 & 58.8 & 2.56 & 1334 & 66 & 1168 & 100\\
SA 82 & 6.3   & 66.3 & 1.20 & 909  & 79 & 747  & 83\\
\hline
\end{tabular}
\end{table}

\begin{figure*}
\resizebox{17cm}{10cm}{\includegraphics*{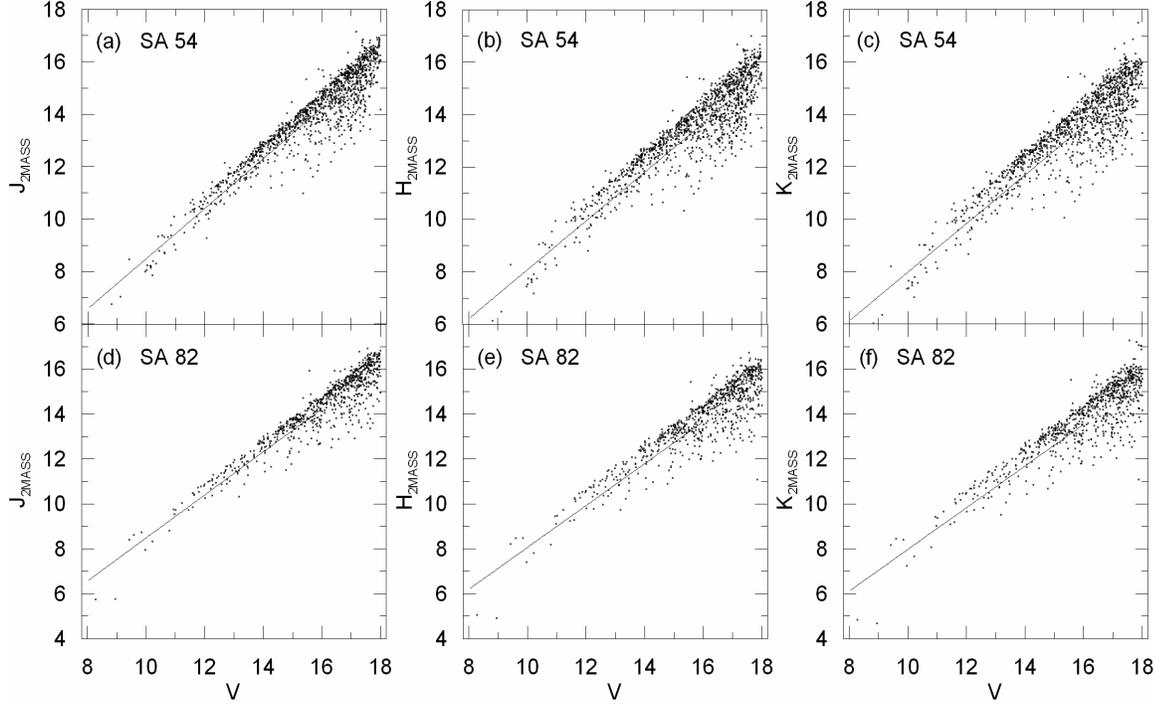}} 
\caption {Two magnitude diagrams for stars in two fields: (a), (b) and (c) for 
SA 54; (d), (e) and (f) for SA 82. The solid line is adopted from Fig. 1.}
\end {figure*}

\begin{figure}
\resizebox{8cm}{6cm}{\includegraphics*{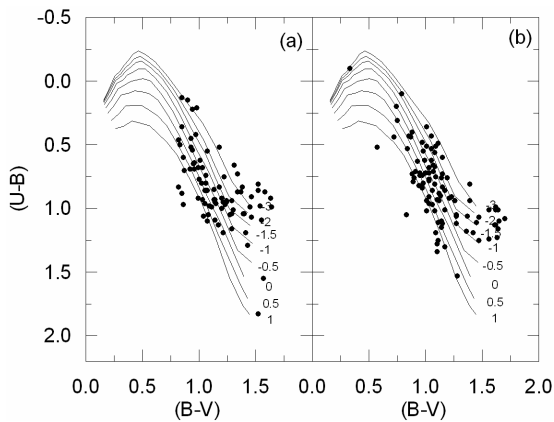}} 
\caption{($U-B, B-V$) two-colour diagrams for two fields. (a) for SA 54 and (b) for 
SA 82. The iso-metallicity lines are taken from Lejuene et al. (1997).}
\end {figure}

\section{Application of the procedure to the giants in two star fields}
We applied the procedure cited above to separate the giants and dwarfs in two 
high latitude fields, SA 54 and SA 82, investigated by Basle astronomers 
(Becker et al. 1982, 1991) in the UBV photometry. Data for the fields are given in
Table 2. We excluded the extragalactic objects via the procedure of Bilir, 
Karata\c{s} \& Ak (2003), then $J/V$, $H/V$, and $K/V$ two magnitude diagrams 
were drawn for stars in each field (Fig. 2). The distribution of stars about 
a central line with small deviations is a strong evidence for the matching 
procedure between spectroscopic catalogues and 2MASS catalogue. Although all the 
2MASS magnitudes are available for dwarf-giant separation we preferred the $J$ 
magnitude. The limiting apparent magnitude is $V=18$ for both fields, 
however we restricted our work with $V=16$, where the procedure in Section 2 
is defined. The number of giants in SA 54 and SA 82 turned out to be 100 and 
83, respectively. They are plotted in Fig. 3 and their metal abundances,  
necessary for absolute magnitude determination, are evaluated by means 
of iso-metallicity lines of Lejuene, Cuisinier and Buser (1997) which are 
available for $\log g=3$ and cover an interval of $-3\leq[Fe/H]\leq+1$ 
dex. The absolute magnitudes of 183 giants in two fields are determined by 
the following equations of Ratnatunga \& Freeman (1989):

\begin{equation}
M(V)=A_{0}+1.11[Fe/H]+0.443	
\end{equation}
\begin{equation}
A_{0}=68.08+170.3C+163.4C^{2}-71.20C^{3}+11.75C^{4}
\end{equation}
\begin{equation}
C=min \{(B-V)_{0}-0.10[Fe/H]-0.013, 1.80\}
\end{equation}
The distance to a star relative to the Sun is carried out by the following 
formula:
\begin{equation}
[V-M(V)]_{0}=5\log r-5								
\end{equation}
Then, the vertical distance to the Galactic plane ($z$) of a star could be 
evaluated by its distance $r$ and its Galactic latitude ($b$) which could be 
provided by its right ascension and declination. 

Table 3 and Table 4 give the logarithmic space density functions, $D^{*}=\log 
D+10$, for SA 54 and SA 82 respectively, where $D=N/ \Delta V_{1,2}$; 
$\Delta V_{1,2}=(\pi/180)^{2}(\sq/3)(r_{2}^{3}-r_{1}^{3})$; $\sq$ denotes the 
size of the field; $r_{1}$ and $r_{2}$ denote the limiting distance of the volume 
$\Delta V_{1,2}$; and $N$ denotes the number of stars.

\setcounter{table}{2}
\begin{table}
\center
\caption{The logarithmic space density function $D^{*}=\log D+10$ for giants in SA 54. 
$z^{*}$ is the mean distance from the Galactic plane and a figure in the bracket, in 
the column of volume, show that the value at the left of it will be multiplied by ten 
to the power of this figure. The other symbols are explained in the text (distances in 
kpc, volumes in pc$^{3}$).}
\begin{tabular}{ccccc}
\hline
$r_{1}-r_{2}$ & $\Delta V_{1,2}$ & N   & $z^{*}$ & $D^{*}$\\
\hline
0 - 1 & 2.60~(05) & 24 & 0.52 & 5.97 \\
1 - 2 & 1.82~(06) & 17 & 1.34 & 4.97 \\
2 - 5 & 3.04~(07) & 13 & 2.86 & 3.63 \\ 
5 -10 & 2.27~(08) & 14 & 6.14 & 2.79 \\
10-15 & 6.17~(08) &  8 & 10.78& 2.11 \\
15-25 & 3.18~(09) & 10 & 18.18& 1.50 \\
25-40 & 1.26~(10) &  9 & 27.03& 0.85 \\
40-54 & 2.43~(10) &  5 & 43.78& 0.31 \\
\hline
\end{tabular}
\end{table}

\setcounter{table}{3}
\begin{table}
\center
\caption{The logarithmic space density function $D^{*}=\log D+10$ for giants in SA 82
(symbols as in Table 3).}
\begin{tabular}{ccccc}
\hline
$r_{1}-r_{2}$ & $\Delta V_{1,2}$ & N   & $z^{*}$ & $D^{*}$\\
\hline
0 - 1  & 1.22~(05) &  7 & 0.56 & 5.76 \\
1 - 2  & 8.53~(05) & 10 & 1.15 & 5.07 \\
2 - 5  & 1.43~(07) & 16 & 3.04 & 4.05 \\ 
5- 7.5 & 3.62~(07) &  9 & 5.56 & 3.40 \\
7.5-15 & 3.60~(08) &  9 & 11.47& 2.40 \\
15-25  & 1.49~(09) &  9 & 18.28& 1.78 \\
25-40  & 5.89~(09) &  8 & 27.83& 1.13 \\
40-60  & 1.85~(10) &  6 & 45.47& 0.51 \\
60-90  & 6.25~(10) &  9 & 70.31& 0.16 \\
\hline
\end{tabular}
\end{table}

\setcounter{table}{4}
\begin{table*}
\center
\caption{Galactic model parameters, for two fields, for three populations, 
i.e. thin disc, thick disc, and halo, resulting from the comparison of observed 
and combined space densities with the density law combined for the thin and 
thick discs, and halo. The symbols: $n^{*}$: the logarithmic local space 
density, H: the scaleheight, $\kappa$: the axes ratio for halo, $n^{*}_{tot}$: 
the combined logarithmic local space density for three populations, $\chi^{2}$: 
the statistics used for comparison the observed data with the density law.}
\begin{tabular}{ccccccccc}
\hline
&\multicolumn{2}{c}{Thin disc}& \multicolumn{2}{c}{Thick disc}&\multicolumn{2}{c}{Halo} & &\\
Field& n$^{*}$ & H (pc) & n$^{*}$ & H (pc) & n$^{*}$ & $\kappa$ & n$^{*}_{tot}$ & $\chi^{2}$ ($10^{-10}$)\\
\hline  
SA 54 & $6.64_{-0.01}^{+0.01}$ & $301_{-3}^{+3}$ & $5.58_{-0.03}^{+0.03}$ & $583_{-22}^{+23}$ & $3.66_{-0.15}^{+0.12}$ & $0.74_{-0.21}^{+0.26}$ & 6.68 & 178\\
SA 82 & $6.60_{-0.01}^{+0.01}$ & $259_{-3}^{+4}$ & $5.34_{-0.04}^{+0.03}$ & $927_{-39}^{+43}$ & $3.61_{-0.10}^{+0.07}$ & $0.74_{-0.09}^{+0.07}$ & 6.62 & 341\\
\hline
\end{tabular}
\end{table*}

\section{Galactic model parameters} 
The comparison of the observed space density functions given in Tables 3 and 4 
and the density law combined for three populations, i.e. thin disc, thick disc 
and halo is carried out in Fig. 4, by $\chi^{2}$ method. The density 
laws for three populations are not given here, however one can find them in many 
papers (cf. Karaali et al. 2004a). The resultant Galactic model parameters for 
three populations and the total local logarithmic space density for giants, for 
two fields, are given in Table 5. The most conspicuous numerical values are for the 
total local densities for giants in two fields, i.e. $n^{*}=6.68$ and $n^{*}=6.62$ 
for SA 54 and SA 82, respectively, which are rather close to the one of Gliese 
(1969), i.e. $\odot=6.64$. Although the axes ratio for the halo components of two 
fields are equal to each other, $\kappa=0.74$, there are meaningful differences 
between the corresponding parameters for two fields, such as the scaleheights, 
for example. The scaleheight for the thick disc for SA 54 is 583 pc, whereas the 
one for SA 82 is 927 pc, which confirms our suggestion that the Galactic model 
parameters are Galactic longitude dependent (Bilir et al. 2005b). Actually, the 
Galactic latitudes of these fields are close to each other, $b=+58^{o}.8$ of 
SA 54 and $b=+66^{o}.3$ of SA 82, whereas SA 54 is almost in the anticenter 
direction ($l=200^{o}.1$) and SA 82 is in the center direction ($l=6^{o}.3$) 
of the Galaxy (for detail see Section 5).

\begin{figure}
\resizebox{8cm}{9.78cm}{\includegraphics*{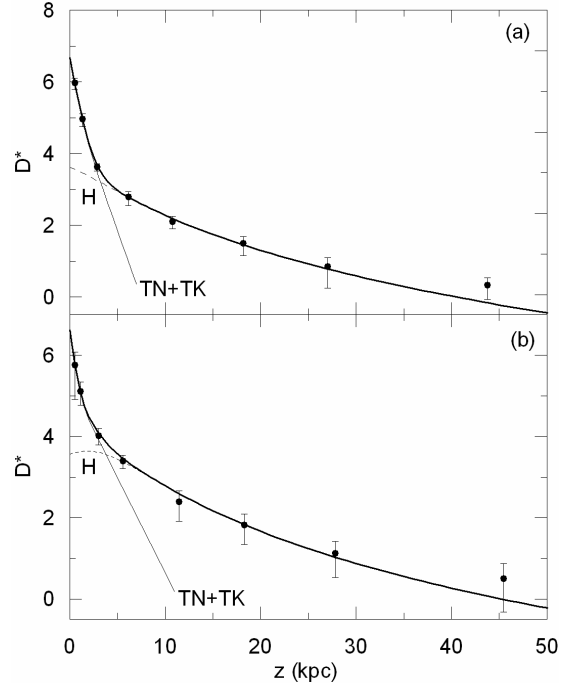}} 
\caption {Comparison of the observed and combined space density function 
($\bullet$) for the thin and thick discs and halo with the combined density 
law, for giants in two fields: (a) for SA 54 and (b) for SA 82. The thin 
and dashed lines show the density laws of two discs (TN+TK) and Halo (H) 
respectively.}
\end {figure}

\section{Conclusion}

We presented a different procedure for separation of field dwarfs and field 
giants by comparison of their 2MASS and V magnitudes down to the limiting 
magnitude of $V=16$. The procedure is based on the spectroscopically selected 
standards, hence it is confident. We applied this procedure to stars in two 
fields, SA 54 and SA 82, and we estimated the Galactic model parameters for 
thin disc, thick disc and halo giants as well as their total local space 
density (Table 5). 

Our results are in agreement with the ones appeared in the literature so far, 
however the model parameters estimated by many researchers are restricted to 
only scaleheight of discs. Whereas in this work, we give a full set of model 
parameters for three populations, thin disc, thick disc and halo. The pioneer 
works give rough scaleheights for discs (thin and thick discs), and claim for 
the existence of halo giants. For example, Pritchet (1983), Bahcall \& Soneira 
(1984) and Buser \& Kaeser (1985) give 150-250 pc, 250$\pm$100 pc, and less 
than 300 pc for disc scaleheight, respectively. Mendez \& van Altena (1996) 
give more precise values, i.e. a mean scaleheight of 250$\pm$32 pc for disc 
subgiants and they claim that it is in agreement with previous scaleheight for 
red giants. The most recent scaleheights for thin and thick disc giants given 
by Cabrera-Lavers et al. (2005) are in agreement with the ones 
estimated in our work for SA 82 giants. Actually, they give 268.81$\pm$12.65 pc 
and 1061.9$\pm$52.16 pc, for thin and thick disc giants which are close to 259 
pc and 927 pc given in Table 5. The agreement of the (total) local logarithmic 
space densities estimated for giants in two fields, $D^{*}(0)=6.68$ and  
$D^{*}(0)=6.62$ for SA 54 and SA 82 respectively, with the ones of Gliese (1969), 
$\odot=6.64$, is a strong confirmation both for the Galactic model parameters 
for giants and for the procedure used for separation of field 
dwarfs and field giants. 

These new calibrations and procedure having been adapted into robotic 
telescopes, i.e. ROTSE I-III, can be applied efficiently in separation 
of the dwarfs and giants stars in the star fields. The giant candidates 
selected through this procedure can be used in giant stars survey as in Grid 
Giant Star Survey (Bizyaev et al. 2005). Also we applied this method to 
NGC 1513 open cluster using ROTSE-IIId (Akerlof et al. 2003) telescope located 
in Bak\i rl\i tepe Antalya, and we could separate the dwarfs and giants easily.
Hence, we can say that this procedure would be used as a practical tool in 
separation of giants and dwarfs in the sky surveys carried out by robotic 
telescope in red bands.  
  
Finally, we should emphasize that there are systematic differences between the 
model parameters for giants in two fields which are investigated homogeneously.
Although the galactic latitudes of these fields are close to each other, 
$b=+58^{o}.8$ and $b=+66^{o}.3$ for SA 54 and SA 82, respectively, their 
galactic longitudes are quite different, $l=200^{o}.1$ and $l=6^{o}.3$ for the 
fields in the same order. Hence, it seems that the systematic differences 
between the model parameters for giants in two fields originate from their 
difference in longitude which confirms our suggestion that Galactic model 
parameters are Galactic longitude dependent (Bilir et al. 2005b).

\begin{acknowledgements}
We would like to thank the anonymous referee for insightful comments and 
suggestions that improved this paper greatly. We also thank H. \c{C}akmak for 
preparing some computer software for this study. This research has made use of 
the SIMBAD database, operated at CDS, Strasbourg, France. This publication makes 
use of data products from the Two Micron All Sky Survey, which is a joint 
project of the University of Massachusetts and the Infrared Processing and 
Analysis Center/California Institute of Technology, funded by the National 
Aeronautics and Space Administration and the National Science Foundation.

\end{acknowledgements}

\setcounter{table}{0}
\begin{table*}
{\tiny
\center
\caption{V and 2MASS magnitudes of Cayrel et al. (2001) and Ratnatunga \& 
Freeman (1989) for stars in our sample. The coordinates are for the in epoch 2000. 
Symbols: (g) and (d): giant and dwarf respectively; (1) and (2) correspond to the 
references of Cayrel et al. (2001), and Ratnatunga \& Freeman (1989), respectively.}
\begin{tabular}{lcccrrrrrrcr}
\hline
Star/field &  $\alpha$ & $\delta$ &     V & \multicolumn{2} {c} {J} & \multicolumn{2} {c} {H} & \multicolumn{2} {c} {K}&       type &       refs    \\
 &  (h~~m~~s) & ($^{o}~^{'}~^{''}$) &     (mag) & \multicolumn{2} {c} {(mag)} & \multicolumn{2} {c} {(mag)} & \multicolumn{2} {c} {(mag)}&    &     \\
\hline
 HD 000026 &   00 05 22 &  +08 47 16 & 8.22 &      6.540 & $\pm$0.019 &      6.106 & $\pm$0.020 &      6.032 & $\pm$0.017 &          g &          1 \\
 HD 000097 &   00 05 46 &  -19 40 13 & 9.62 &      8.057 &      0.019 &      7.612 &      0.034 &      7.513 &      0.023 &          g &          1 \\
BPS CS 22876-0032 &   00 07 37 &  -35 31 17 &      12.84 &     11.802 &      0.022 &     11.555 &      0.024 &     11.485 &      0.021 &  d  &    1 \\
BD +13 0013 &   00 12 30 & +14 33 49 & 8.59 &      6.847 &      0.023 &      6.380 &      0.018 &      6.245 &      0.020 &          g &          1 \\
 HD 000987 &   00 13 53 &  -74 41 18 & 8.74 &      7.406 &      0.021 &      7.087 &      0.029 &      6.962 &      0.023 &          d &          1 \\
CD -23 00072 &   00 16 17 & -22 34 41 &9.58 &      8.147 &      0.021 &      7.786 &      0.046 &      7.683 &      0.021 &          g &          1 \\
BPS CS 29527-0015 &   00 29 06 &  -19 07 00 &      14.24 &     13.298 &      0.028 &     13.080 &      0.036 &     13.048 &      0.036 &  d &     1 \\
 HD 002796 &   00 31 17 &  -16 47 41 & 8.51 &      6.853 &      0.020 &      6.391 &      0.034 &      6.256 &      0.016 &          g &          1 \\
 HD 003008 &   00 33 14 &  -10 43 43 & 9.54 &      7.181 &      0.018 &      6.603 &      0.034 &      6.417 &      0.017 &          g &          1 \\
BD +21 0064 &   00 34 16 &  +22 24 48& 9.65 &      7.634 &      0.024 &      7.135 &      0.024 &      6.916 &      0.017 &          g &          1 \\
 HD 003567 &   00 38 32 &  -08 18 33 & 9.26 &      8.218 &      0.019 &      7.933 &      0.034 &      7.889 &      0.026 &          d &          1 \\
BPS CS 29497-0034 &   00 41 40 &  -26 18 55 &      13.30 &     10.571 &      0.024 &     10.115 &      0.024 &      9.996 &      0.021 & g &      1 \\
 HD 004306 &   00 45 27 &  -09 32 40 & 9.08 &      7.424 &      0.026 &      6.975 &      0.069 &      6.823 &      0.023 &          g &          1 \\
CD -38 00245&  00 46 36 &  -37 39 34 &12.00 &     10.279 &      0.024 &      9.762 &      0.026 &      9.648 &      0.024 &          g &          1 \\
 HD 004893 &   00 49 23 &  -73 28 43 &8.47  &      6.013 &      0.019 &      5.348 &      0.033 &      5.118 &      0.018 &          g &          1 \\
 HD 005426 &   00 55 41 &  -33 45 11 &9.63  &      8.082 &      0.034 &      7.626 &      0.031 &      7.548 &      0.027 &          g &          1 \\
    SA 141 &   00 58 32 &  -29 48 22 &13.98 &     12.004 &      0.028 &     11.326 &      0.022 &     11.248 &      0.019 &          g &          2 \\
    SA 141 &   00 59 08 &  -29 40 22 &14.04 &     12.223 &      0.021 &     11.690 &      0.022 &     11.618 &      0.021 &          g &          2 \\
    SA 141 &   00 59 17 &  -30 16 03 &14.57 &     11.504 &      0.027 &     10.753 &      0.027 &     10.555 &      0.027 &          g &          2 \\
    SA 141 &   00 59 34 &  -29 37 30 &14.38 &     12.355 &      0.024 &     11.779 &      0.025 &     11.680 &      0.023 &          g &          2 \\
    SA 141 &   01 00 29 &  -31 13 18 &15.22 &     13.373 &      0.030 &     12.853 &      0.026 &     12.711 &      0.027 &          g &          2 \\
    SA 141 &   01 00 29 &  -32 01 30 &15.71 &     13.702 &      0.029 &     13.154 &      0.023 &     13.042 &      0.030 &          g &          2 \\
    SA 141 &   01 00 32 &  -29 26 32 &15.27 &     13.394 &      0.022 &     12.813 &      0.021 &     12.699 &      0.024 &          g &          2 \\
    SA 141 &   01 01 09 &  -28 15 26 &14.38 &     12.320 &      0.023 &     11.686 &      0.022 &     11.530 &      0.019 &          g &          2 \\
    SA 141 &   01 01 11 &  -30 27 55 &15.08 &     12.906 &      0.026 &     12.207 &      0.024 &     12.116 &      0.026 &          g &          2 \\
    SA 141 &   01 02 55 &  -30 12 36 &14.40 &     12.034 &      0.024 &     11.372 &      0.027 &     11.238 &      0.021 &          g &          2 \\
 HD 006254 &   01 03 16 &  -26 10 48 & 8.01 &      6.203 &      0.018 &      5.675 &      0.033 &      5.559 &      0.020 &          g &          1 \\
 HD 006268 &   01 03 18 &  -27 52 50 & 8.10 &      6.335 &      0.020 &      5.843 &      0.023 &      5.714 &      0.018 &          g &          1 \\
    SA 141 &   01 03 34 &  -31 02 40 &13.22 &     11.301 &      0.026 &     10.725 &      0.025 &     10.611 &      0.021 &          g &          2 \\
 HD 006229 &   01 03 36 &  +23 46 06 &8.60  &      7.088 &      0.030 &      6.646 &      0.026 &      6.575 &      0.017 &          g &          1 \\
    SA 141 &   01 04 25 &  -30 01 52 &15.59 &     13.743 &      0.026 &     13.134 &      0.022 &     13.003 &      0.027 &          g &          2 \\
    SA 141 &   01 05 16 &  -30 01 56 &13.64 &     11.850 &      0.022 &     11.298 &      0.023 &     11.237 &      0.029 &          g &          2 \\
BPS CS 22166-0030&01 05 28& -11 57 31&13.58 &     12.484 &      0.024 &     12.149 &      0.023 &     12.099 &      0.021 &          d &          1 \\
 HD 006834 &   01 09 35 &  +39 46 52 &8.42  &      7.520 &      0.023 &      7.317 &      0.020 &      7.286 &      0.016 &          d &          1 \\
 HD 007041 &   01 09 49 &  -56 21 22 &9.04  &      7.459 &      0.021 &      6.993 &      0.049 &      6.901 &      0.021 &          g &          1 \\
    SA 141 &   01 09 60 &  -29 28 47 &14.49 &     12.486 &      0.027 &     11.905 &      0.023 &     11.757 &      0.021 &          g &          2 \\
    SA 141 &   01 10 11 &  -29 25 29 &14.05 &     12.341 &      0.026 &     11.817 &      0.023 &     11.729 &      0.021 &          g &          2 \\
 HD 007172 &   01 11 43 &  -25 57 31 &8.79  &      8.580 &      0.030 &      8.277 &      0.053 &      8.181 &      0.026 &          d &          1 \\
    SA 141 &   01 11 53 &  -31 23 06 &13.34 &     11.476 &      0.021 &     10.946 &      0.024 &     10.816 &      0.021 &          g &          2 \\
    SA 141 &   01 11 55 &  -30 26 34 &15.70 &     13.657 &      0.024 &     13.077 &      0.028 &     12.948 &      0.033 &          g &          2 \\
    SA 141 &   01 12 21 &  -30 02 11 &13.32 &     11.065 &      0.024 &     10.313 &      0.022 &     10.230 &      0.019 &          g &          2 \\
    SA 141 &   01 13 23 &  -31 56 10 &13.38 &     11.686 &      0.023 &     11.181 &      0.022 &     11.115 &      0.023 &          g &          2 \\
    SA 141 &   01 13 29 &  -31 16 01 &14.19 &     12.478 &      0.023 &     11.891 &      0.023 &     11.826 &      0.023 &          g &          2 \\
 HD 007424 &   01 14 07 &  -16 25 35 &10.08 &      8.859 &      0.018 &      8.532 &      0.044 &      8.460 &      0.024 &          d &          1 \\
    SA 141 &   01 14 26 &  -29 35 18 &15.52 &     13.497 &      0.024 &     12.912 &      0.026 &     12.722 &      0.027 &          g &          2 \\
    SA 141 &   01 15 19 &  -30 48 12 &13.24 &     11.056 &      0.023 &     10.457 &      0.023 &     10.303 &      0.023 &          g &          2 \\
    SA 141 &   01 15 25 &  -30 01 59 &13.64 &     11.820 &      0.024 &     11.258 &      0.023 &     11.155 &      0.024 &          g &          2 \\
 HD 007595 &   01 15 30 &  -28 15 54 & 9.73 &      7.601 &      0.026 &      6.963 &      0.044 &      6.810 &      0.017 &          g &          1 \\
    SA 141 &   01 15 44 &  -27 58 53 &15.72 &     13.597 &      0.030 &     12.965 &      0.026 &     12.871 &      0.033 &          g &          2 \\
    SA 141 &   01 18 49 &  -31 46 01 &15.96 &     14.184 &      0.030 &     13.618 &      0.040 &     13.511 &      0.042 &          g &          2 \\
 HD 007983 &   01 18 60 &  -08 56 22 & 8.89 &      7.718 &      0.027 &      7.412 &      0.036 &      7.354 &      0.021 &          d &          1 \\
    SA 141 &   01 19 26 &  -30 31 55 &15.69 &     13.652 &      0.024 &     13.073 &      0.031 &     12.977 &      0.034 &          g &          2 \\
    SA 141 &   01 19 33 &  -29 31 07 &13.93 &     11.617 &      0.027 &     10.967 &      0.023 &     10.847 &      0.021 &          g &          2 \\
BPS CS 22953-0037&01 25 07&-59 15 57 &13.64 &     12.683 &      0.026 &     12.444 &      0.028 &     12.461 &      0.034 &          d &          1 \\
 HD 008724 &   01 26 18 &  +17 07 35 & 8.34 &      6.299 &      0.023 &      5.769 &      0.040 &      5.636 &      0.020 &          g &          1 \\
 G 002-038 &   01 26 56 &  +12 00 54 &11.40 &     10.257 &      0.026 &      9.972 &      0.024 &      9.906 &      0.021 &          d &          1 \\
 HD 009051 &   01 28 47 &  -24 20 25 & 8.92 &      7.289 &      0.020 &      6.831 &      0.047 &      6.707 &      0.016 &          g &          1 \\
 HD 009430 &   01 32 58 &  +23 41 44 & 9.03 &      7.780 &      0.018 &      7.471 &      0.017 &      7.401 &      0.020 &          d &          1 \\
BD -18 0255 &  01 33 17 &  -18 12 40 &10.27 &      9.416 &      0.022 &      9.247 &      0.022 &      9.237 &      0.023 &          d &          1 \\
CD -61 00282&  01 36 06 &  -61 05 03 &10.10 &      8.948 &      0.039 &      8.649 &      0.034 &      8.682 &      0.027 &          d &          1 \\
 HD 009902 &   01 37 09 &  +20 42 00 &8.71  &      7.308 &      0.030 &      6.906 &      0.024 &      6.839 &      0.020 &          d &          1 \\
BD -18 0271 &   01 37 19 &  -17 29 04&9.85  &      7.530 &      0.021 &      6.914 &      0.034 &      6.786 &      0.024 &          g &          1 \\
 HD 011377 &   01 51 29 &  -16 44 26 &8.50  &      7.532 &      0.026 &      7.318 &      0.036 &      7.218 &      0.027 &          d &          1 \\
 HD 011397 &   01 51 41 &  -16 19 04 & 9.02 &      7.686 &      0.023 &      7.318 &      0.033 &      7.270 &      0.023 &          d &          1 \\
BD +29 0366 &   02 10 25 &  +29 48 24 &8.77 &      7.572 &      0.018 &      7.280 &      0.026 &      7.216 &      0.015 &          d &          1 \\
BD -01 0306 &   02 14 40 &  -01 12 05 &9.09 &      7.899 &      0.021 &      7.589 &      0.036 &      7.520 &      0.026 &          d &          1 \\
 HD 013979 &   02 15 21 &  -25 54 55 & 9.19 &      7.645 &      0.027 &      7.203 &      0.059 &      7.093 &      0.021 &          g &          1 \\
 HD 014056 &   02 17 07 &  +21 34 01 & 9.04 &      7.853 &      0.030 &      7.521 &      0.021 &      7.466 &      0.021 &          d &          1 \\
BD +13 0374 &   02 20 39 &  +13 40 12& 8.99 &      7.672 &      0.032 &      7.279 &      0.031 &      7.171 &      0.021 &          d &          1 \\
 HD 015306 &   02 27 51 &  +00 47 22 & 8.92 &      8.182 &      0.021 &      8.013 &      0.033 &      7.980 &      0.023 &          d &          1 \\
 HD 016031 &   02 34 11 &  -12 23 03 & 9.78 &      8.790 &      0.027 &      8.516 &      0.034 &      8.457 &      0.019 &          d &          1 \\
 HD 016115 &   02 35 06 &  -09 26 34 & 8.15 &      6.062 &      0.023 &      5.544 &      0.018 &      5.322 &      0.023 &          g &          1 \\
BD +09 0352 &   02 41 14 &  +09 46 12&10.17 &      9.012 &      0.027 &      8.829 &      0.025 &      8.749 &      0.021 &          d &          1 \\
BPS CS 22189-0009&02 41 42& -13 28 15&14.04 &     12.391 &      0.021 &     11.895 &      0.025 &     11.798 &      0.028 &          g &          1 \\
 G 004-036 &   02 43 22 &  +13 25 57 &11.49 &     10.362 &      0.021 &     10.058 &      0.021 &     10.012 &      0.019 &          d &          1 \\
 HD 017288 &   02 44 10 &  -60 03 22 & 9.88 &      8.697 &      0.027 &      8.423 &      0.031 &      8.329 &      0.036 &          d &          1 \\
 HD 017548 &   02 48 52 &  -01 30 35 & 8.16 &      7.097 &      0.021 &      6.802 &      0.051 &      6.765 &      0.024 &          d &          1 \\
    SA 189 &   02 50 00 &  -61 52 31 &15.97 &     13.894 &      0.024 &     13.324 &      0.022 &     13.203 &      0.033 &          g &          2 \\
 HD 017820 &   02 51 58 &  +11 22 12 & 8.38 &      7.233 &      0.039 &      6.930 &      0.049 &      6.895 &      0.020 &          d &          1 \\
    SA 189 &   02 54 20 &  -60 02 46 &15.56 &     13.363 &      0.026 &     12.670 &      0.027 &     12.543 &      0.035 &          g &          2 \\
    SA 189 &   02 56 02 &  -58 09 31 &15.74 &     13.957 &      0.028 &     13.414 &      0.023 &     13.298 &      0.032 &          g &          2 \\
    SA 189 &   02 57 30 &  -60 46 40 &13.94 &     11.980 &      0.021 &     11.380 &      0.025 &     11.309 &      0.025 &          g &          2 \\
    SA 189 &   02 59 59 &  -60 25 32 &13.73 &     11.561 &      0.022 &     10.914 &      0.021 &     10.851 &      0.026 &          g &          2 \\
    SA 189 &   03 01 28 &  -59 47 25 &13.56 &     11.739 &      0.021 &     11.188 &      0.023 &     11.106 &      0.023 &          g &          2 \\
    SA 189 &   03 01 55 &  -61 33 00 &15.06 &     13.019 &      0.029 &     12.303 &      0.027 &     12.238 &      0.021 &          g &          2 \\
LP 0831-0070 &   03 06 05& -22 19 18 &11.40 &     10.654 &      0.021 &     10.379 &      0.023 &     10.344 &      0.027 &          d &          1 \\
BPS CS 22968-0014&03 06 29&-54 30 32 &13.72 &     12.026 &      0.024 &     11.533 &      0.026 &     11.467 &      0.025 &          g &          1 \\
 HD 019745 &   03 07 03 &  -65 26 57 & 9.11 &      7.213 &      0.024 &      6.675 &      0.027 &      6.536 &      0.020 &          g &          1 \\
    SA 189 &   03 07 32 &  -58 14 48 &15.16 &     13.274 &      0.026 &     12.669 &      0.023 &     12.599 &      0.029 &          g &          2 \\
 HD 019445 &   03 08 26 &  +26 19 51 & 8.05 &      6.948 &      0.020 &      6.696 &      0.034 &      6.640 &      0.020 &          d &          1 \\
 HD 020038 &   03 10 27 &  -58 49 41 & 8.93 &      7.332 &      0.023 &      6.850 &      0.029 &      6.780 &      0.021 &          g &          1 \\
    SA 189 &   03 11 17 &  -59 16 44 &13.66 &     11.829 &      0.023 &     11.285 &      0.022 &     11.193 &      0.021 &          g &          2 \\
BPS CS 22172-0002&03 14 21&-10 35 10 &12.73 &     10.898 &      0.022 &     10.431 &      0.022 &     10.320 &      0.021 &          g &          1 \\
    SA 189 &   03 19 13 &  -58 09 15 &13.41 &     11.539 &      0.024 &     10.999 &      0.023 &     10.863 &      0.026 &          g &          2 \\
    SA 189 &   03 19 23 &  -58 45 50 &15.03 &     13.099 &      0.024 &     12.483 &      0.025 &     12.392 &      0.027 &          g &          2 \\
CD -33 01173 & 03 19 35 &  -32 50 43 &10.91 &     10.008 &      0.022 &      9.790 &      0.025 &      9.745 &      0.023 &          d &          1 \\
    SA 189 &   03 21 10 &  -61 08 06 &13.42 &     11.139 &      0.022 &     10.506 &      0.022 &     10.369 &      0.023 &          g &          2 \\
    SA 189 &   03 21 31 &  -58 26 06 &14.78 &     13.020 &      0.023 &     12.495 &      0.024 &     12.405 &      0.024 &          g &          2 \\
    SA 189 &   03 22 53 &  -60 58 22 &14.89 &     13.045 &      0.024 &     12.533 &      0.023 &     12.397 &      0.025 &          g &          2 \\
    SA 189 &   03 23 47 &  -60 17 03 &15.15 &     13.085 &      0.030 &     12.469 &      0.025 &     12.381 &      0.019 &          g &          2 \\
    SA 189 &   03 27 30 &  -60 06 25 &15.30 &     13.214 &      0.024 &     12.638 &      0.025 &     12.520 &      0.024 &          g &          2 \\
BD +20 0571 &   03 27 40 & +21 02 48 &10.79 &      9.495 &      0.018 &      9.163 &      0.016 &      9.131 &      0.019 &          d &          1 \\
 HD 021543 &   03 28 21 &  -06 31 51 & 8.23 &      7.001 &      0.018 &      6.743 &      0.040 &      6.647 &      0.018 &          d &          1 \\
 HD 021581 &   03 28 54 &  +00 25 03 & 8.72 &      6.975 &      0.018 &      6.547 &      0.047 &      6.414 &      0.020 &          g &          1 \\
CD -47 01087 & 03 34 44 &  -47 16 12 &10.26 &      9.105 &      0.022 &      8.791 &      0.024 &      8.726 &      0.023 &          d &          1 \\
 HD 022670 &   03 39 04 &  +09 39 10 & 9.01 &      7.898 &      0.023 &      7.623 &      0.029 &      7.583 &      0.031 &          d &          1 \\
 HD 022785 &   03 39 38 &  -04 08 54 & 9.59 &      8.746 &      0.023 &      8.552 &      0.038 &      8.525 &      0.026 &          d &          1 \\
BPS CS 29529-0012&03 43 26&-60 20 20 &14.80 &     13.562 &      0.024 &     13.272 &      0.034 &     13.274 &      0.034 &          d &          1 \\
 HD 023798 &   03 46 46 &  -30 51 13 & 8.32 &      6.276 &      0.030 &      5.657 &      0.036 &      5.517 &      0.024 &          g &          1 \\
 HD 024339 &   03 51 24 &  -25 55 57 & 9.44 &      8.303 &      0.026 &      8.035 &      0.029 &      7.970 &      0.024 &          d &          1 \\
BPS CS 29529-0034&03 53 53&-59 06 27 &14.94 &     13.612 &      0.027 &     12.967 &      0.035 &     12.732 &      0.029 &          d &          1 \\
 HD 025169 &   03 57 52 &  -46 22 59 & 8.03 &      7.035 &      0.024 &      6.810 &      0.033 &      6.733 &      0.018 &          d &          1 \\
 HD 026169 &   04 00 52 &  -75 36 11 & 8.79 &      7.172 &      0.029 &      6.695 &      0.040 &      6.634 &      0.031 &          g &          1 \\
 HD 025704 &   04 01 45 &  -57 12 25 & 8.10 &      6.977 &      0.026 &      6.650 &      0.020 &      6.556 &      0.029 &          d &          1 \\
\end{tabular}  
}  
\end{table*}

\begin{table*}
{\tiny
\begin{tabular}{lcccrrrrrrcr}
\hline
Star/field &  $\alpha$ & $\delta$ &     V & \multicolumn{2} {c} {J} & \multicolumn{2} {c} {H} & \multicolumn{2} {c} {K}&       type &       refs \\
&  (h~~m~~s) & ($^{o}~^{'}~^{''}$) &     (mag) & \multicolumn{2} {c} {(mag)} & \multicolumn{2} {c} {(mag)} & \multicolumn{2} {c} {(mag)}&    &     \\
\hline
 HD 025532 &   04 04 11 &  +23 24 27 & 8.24 &      6.688 &$\pm$  0.023 &      6.327 & $\pm$0.021 &      7.183 &$\pm$ 0.021 &          g &          1 \\
BD +06 0648 &   04 13 13 &  +06 36 02& 9.09 &      6.562 &      0.020 &      5.905 &      0.033 &      5.798 &      0.018 &          g &          1 \\
 HD 284248 &   04 14 36 &  +22 21 04 & 9.22 &      8.165 &      0.018 &      7.927 &      0.023 &      7.871 &      0.023 &          d &          1 \\
BD -06 0855 &   04 14 58 &  -05 37 49&10.60 &      9.332 &      0.029 &      8.888 &      0.029 &      8.793 &      0.023 &          d &          1 \\
BPS CS 22182-0024&04 20 01& -31 21 51&12.88 &     11.952 &      0.026 &     11.740 &      0.023 &     11.682 &      0.024 &          d &          1 \\
BPS CS 22186-0050&04 32 08& -37 03 16&13.64 &     12.598 &      0.027 &     12.276 &      0.025 &     12.271 &      0.021 &          d &          1 \\
BPS CS 22191-0017&04 36 31& -40 14 45&13.86 &     12.833 &      0.023 &     12.591 &      0.022 &     12.517 &      0.029 &          d &          1 \\
CPD -57 0680&     04 36 40& -57 32 41& 9.26 &      7.588 &      0.020 &      7.108 &      0.042 &      6.957 &      0.026 &          g &          1 \\
 HD 029907 &   04 38 22 &  -65 24 58 & 9.91 &      8.512 &      0.021 &      8.192 &      0.053 &      8.090 &      0.020 &          d &          1 \\
 HD 029574 &   04 38 56 &  -13 20 48 & 8.38 &      5.726 &      0.021 &      5.208 &      0.076 &      4.840 &      0.017 &          g &          1 \\
 HD 029528 &   04 39 37 &  +13 07 23 & 8.95 &      7.606 &      0.018 &      7.280 &      0.027 &      7.168 &      0.020 &          d &          1 \\
 HD 030328 &   04 46 45 &  +02 42 02 & 8.82 &      6.982 &      0.020 &      6.465 &      0.024 &      6.377 &      0.017 &          g &          1 \\
 HD 031128 &   04 52 10 &  -27 03 51 & 9.14 &      8.032 &      0.023 &      7.800 &      0.027 &      7.738 &      0.018 &          d &          1 \\
BD +03 0740 &   05 01 17 &  +04 06 37& 9.82 &      8.795 &      0.021 &      8.532 &      0.034 &      8.516 &      0.029 &          d &          1 \\
 HD 034328 &   05 13 05 &  -59 38 44 & 9.46 &      8.316 &      0.021 &      8.046 &      0.029 &      7.998 &      0.026 &          d &          1 \\
 HD 273975 &   05 18 20 &  -48 52 13 &10.72 &      9.514 &      0.029 &      9.167 &      0.022 &      9.095 &      0.021 &          d &          1 \\
 HD 035179 &   05 22 21 &  -14 23 50 & 9.48 &      7.715 &      0.029 &      7.234 &      0.040 &      7.099 &      0.024 &          g &          1 \\
CD -29 02277 &   05 28 51 & -29 53 48&11.61 &     10.487 &      0.026 &     10.216 &      0.022 &     10.143 &      0.023 &          d &          1 \\
 HD 036552 &   05 30 20 &  -43 41 38 & 8.08 &      6.522 &      0.019 &      6.138 &      0.027 &      5.970 &      0.018 &          g &          1 \\
 HD 036702 &   05 31 52 &  -38 33 24 & 8.38 &      6.126 &      0.026 &      5.477 &      0.024 &      5.329 &      0.020 &          g &          1 \\
 HD 274939 &   05 33 18 &  -47 56 14 & 9.45 &      7.856 &      0.023 &      7.441 &      0.044 &      7.299 &      0.024 &          g &          1 \\
 HD 040402 &   05 57 19 &  -27 19 53 & 8.60 &      6.926 &      0.032 &      6.465 &      0.029 &      6.314 &      0.031 &          g &          1 \\
 HD 041667 &   06 05 04 &  -32 59 39 & 8.52 &      6.630 &      0.024 &      6.107 &      0.029 &      5.929 &      0.020 &          g &          1 \\
 HD 043767 &   06 15 40 &  -46 07 55 & 8.28 &      8.070 &      0.030 &      8.040 &      0.046 &      8.020 &      0.023 &          d &          1 \\
CD -57 01633 &   07 06 29& -57 27 29 & 9.54 &      8.499 &      0.021 &      8.219 &      0.049 &      8.089 &      0.036 &          d &          1 \\
 HD 053871 &   07 09 27 &  +52 02 19 & 9.07 &      7.992 &      0.021 &      7.769 &      0.024 &      7.711 &      0.020 &          d &          1 \\
 HD 064090 &   07 53 33 &  +30 36 18 & 8.30 &      6.956 &      0.018 &      6.611 &      0.017 &      6.537 &      0.023 &          d &          1 \\
 G 090-036 &   08 00 29 &  +29 00 39 &12.70 &     11.382 &      0.021 &     11.046 &      0.022 &     10.962 &      0.020 &          d &          1 \\
 HD 068988 &   08 18 22 &  +61 27 39 & 8.21 &      7.060 &      0.021 &      6.812 &      0.020 &      6.743 &      0.024 &          d &          1 \\
 HD 233511 &   08 19 23 &  +54 05 10 & 9.71 &      8.617 &      0.021 &      8.386 &      0.017 &      8.329 &      0.015 &          d &          1 \\
 HD 233517 &   08 22 47 &  +53 04 49 & 9.70 &      7.346 &      0.020 &      6.752 &      0.034 &      6.637 &      0.027 &          g &          1 \\
 HD 074462 &   08 48 21 &  +67 26 59 & 8.69 &      6.688 &      0.027 &      6.117 &      0.033 &      6.047 &      0.029 &          g &          1 \\
BD +06 2063 &   08 55 38 &  +06 09 05& 8.98 &      5.299 &      0.037 &      4.446 &      0.198 &      4.192 &      0.055 &          g &          1 \\
BD -05 2678 &   08 59 03 &  -06 23 46&11.92 &     10.610 &      0.023 &     10.244 &      0.023 &     10.187 &      0.021 &          d &          1 \\
 HD 076910 &   08 59 06 &  +00 37 26 & 8.50 &      7.584 &      0.026 &      7.355 &      0.040 &      7.323 &      0.024 &          d &          1 \\
BD -03 2525 &   08 59 10 &  -04 01 37& 9.67 &      8.561 &      0.026 &      8.276 &      0.044 &      8.205 &      0.026 &          d &          1 \\
 G 114-042 &   09 10 45 &  -03 48 12 &12.76 &     11.593 &      0.024 &     11.265 &      0.024 &     11.193 &      0.025 &          d &          1 \\
 G 046-031 &   09 17 04 &  +03 01 30 &10.86 &      9.747 &      0.022 &      9.489 &      0.022 &      9.393 &      0.020 &          d &          1 \\
 HD 079633 &   09 17 08 &  +51 19 20 & 9.04 &      8.393 &      0.023 &      8.306 &      0.029 &      8.235 &      0.021 &          d &          1 \\
 HD 080607 &   09 22 40 &  +50 36 14 & 9.07 &      7.798 &      0.029 &      7.509 &      0.029 &      7.389 &      0.021 &          d &          1 \\
 HD 081713 &   09 26 46 &  -20 38 53 & 8.91 &      7.142 &      0.023 &      6.651 &      0.044 &      6.523 &      0.024 &          g &          1 \\
BD +09 2190 &   09 29 16 &  +08 38 00&11.14 &     10.193 &      0.022 &      9.932 &      0.022 &      9.907 &      0.024 &          d &          1 \\
 HD 083212 &   09 36 20 &  -20 53 15 & 8.34 &      6.307 &      0.021 &      5.729 &      0.034 &      5.609 &      0.023 &          g &          1 \\
 HD 083632 &   09 40 34 &  +26 00 14 & 8.06 &      5.612 &      0.019 &      4.861 &      0.027 &      4.725 &      0.020 &          g &          1 \\
 HD 083769 &   09 40 46 &  +01 01 29 &10.23 &      9.398 &      0.026 &      9.193 &      0.022 &      9.163 &      0.023 &          d &          1 \\
 HD 233666 &   09 42 19 &  +53 28 26 & 9.34 &      7.860 &      0.021 &      7.467 &      0.023 &      7.367 &      0.018 &          g &          1 \\
 HD 083888 &   09 42 47 &  +45 31 02 & 8.84 &      7.980 &      0.020 &      7.814 &      0.017 &      7.732 &      0.017 &          d &          1 \\
 G 116-053 &   09 47 42 &  +33 22 42 &12.98 &     11.788 &      0.021 &     11.483 &      0.016 &     11.411 &      0.017 &          d &          1 \\
BD +09 2242 &   09 48 53 &  +08 58 28& 9.59 &      8.554 &      0.030 &      8.308 &      0.051 &      8.294 &      0.021 &          d &          1 \\
 HD 084937 &   09 48 56 &  +13 44 39 & 8.28 &      7.359 &      0.020 &      7.121 &      0.021 &      7.062 &      0.018 &          d &          1 \\
 HD 237846 &   09 52 39 &  +57 54 59 & 9.98 &      8.355 &      0.026 &      7.931 &      0.049 &      7.816 &      0.018 &          g &          1 \\
 HD 085773 &   09 53 39 &  -22 50 08 & 9.43 &      7.230 &      0.024 &      6.660 &      0.063 &      6.505 &      0.021 &          g &          1 \\
CD -24 08642 &   09 59 35& -24 51 26 & 9.64 &      8.050 &      0.027 &      7.580 &      0.031 &      7.491 &      0.021 &          g &          1 \\
 HD 087140 &   10 04 43 &  +54 20 43 & 9.00 &      7.471 &      0.026 &      7.046 &      0.023 &      6.943 &      0.017 &          g &          1 \\
 HD 087646 &   10 06 41 &  +17 53 42 & 8.07 &      6.897 &      0.021 &      6.612 &      0.021 &      6.536 &      0.018 &          d &          1 \\
 HD 089499 &   10 07 30 &  -85 04 33 & 8.68 &      7.032 &      0.023 &      6.604 &      0.040 &      6.427 &      0.026 &          g &          1 \\
 HD 088609 &   10 14 29 &  +53 33 39 & 8.64 &      6.665 &      0.029 &      6.129 &      0.020 &      6.006 &      0.018 &          g &          1 \\
BD +30 2034 &   10 33 15 &  +29 36 55&10.40 &      8.296 &      0.018 &      7.735 &      0.018 &      7.578 &      0.017 &          g &          1 \\
BD +09 2384 &   10 40 25 &  +08 54 04& 9.85 &      8.242 &      0.037 &      7.753 &      0.055 &      7.648 &      0.029 &          g &          1 \\
 HD 092545 &   10 40 58 &  -12 11 44 & 8.56 &      7.548 &      0.023 &      7.347 &      0.024 &      7.282 &      0.027 &          d &          1 \\
BD +29 2091 &   10 47 23 &  +28 23 56&10.22 &      9.104 &      0.018 &      8.805 &      0.016 &      8.739 &      0.019 &          d &          1 \\
 HD 093529 &   10 47 29 &  -25 26 15 & 9.31 &      7.503 &      0.027 &      6.987 &      0.033 &      6.873 &      0.017 &          g &          1 \\
 HD 093487 &   10 47 56 &  +23 20 07 & 8.71 &      7.225 &      0.018 &      6.856 &      0.023 &      6.764 &      0.021 &          g &          1 \\
 G 196-047 &   10 50 29 &  +56 26 40 &12.58 &     11.282 &      0.022 &     10.987 &      0.032 &     10.831 &      0.020 &          d &          1 \\
 HD 094028 &   10 51 28 &  +20 16 39 & 8.23 &      7.130 &      0.020 &      6.854 &      0.023 &      6.832 &      0.017 &          d &          1 \\
 HD 094956 &   10 57 24 &  -29 16 51 & 8.46 &      6.712 &      0.021 &      6.238 &      0.042 &      6.107 &      0.024 &          g &          1 \\
BD -10 3166 &   10 58 29 &  -10 46 13&10.00 &      8.611 &      0.032 &      8.300 &      0.040 &      8.124 &      0.026 &          d &          1 \\
    SA 127 &   10 59 49 &  -18 36 42 &15.04 &     12.777 &      0.022 &     12.104 &      0.025 &     11.973 &      0.023 &          g &          2 \\
    SA 127 &   10 59 52 &  -19 54 50 &14.73 &     13.010 &      0.023 &     12.513 &      0.021 &     12.391 &      0.023 &          g &          2 \\
    SA 127 &   11 00 15 &  -17 53 56 &15.06 &     13.251 &      0.024 &     12.762 &      0.027 &     12.655 &      0.026 &          g &          2 \\
    SA 127 &   11 00 30 &  -18 +6 26 &14.90 &     12.639 &      0.029 &     11.988 &      0.023 &     11.829 &      0.023 &          g &          2 \\
 HD 095405 &   11 00 34 &  -25 51 23 & 8.33 &      6.431 &      0.030 &      5.848 &      0.055 &      5.718 &      0.029 &          g &          1 \\
    SA 127 &   11 01 01 &  -16 50 19 &15.11 &     12.804 &      0.023 &     12.086 &      0.024 &     11.906 &      0.024 &          g &          2 \\
    SA 127 &   11 01 36 &  -17 48 47 &14.46 &     12.684 &      0.026 &     12.089 &      0.023 &     12.023 &      0.021 &          g &          2 \\
    SA 127 &   11 02 07 &  -18 02 07 &14.23 &     12.280 &      0.026 &     11.700 &      0.026 &     11.572 &      0.023 &          g &          2 \\
    SA 127 &   11 02 16 &  -17 33 16 &14.83 &     12.416 &      0.026 &     11.680 &      0.023 &     11.522 &      0.025 &          g &          2 \\
    SA 127 &   11 02 29 &  -16 09 31 &14.81 &     12.696 &      0.024 &     12.014 &      0.025 &     11.865 &      0.027 &          g &          2 \\
    SA 127 &   11 02 30 &  -16 55 53 &15.35 &     13.656 &      0.024 &     13.126 &      0.025 &     13.075 &      0.038 &          g &          2 \\
    SA 127 &   11 02 44 &  -17 36 00 &15.11 &     13.014 &      0.023 &     12.378 &      0.022 &     12.276 &      0.026 &          g &          2 \\
    SA 127 &   11 03 24 &  -17 14 31 &14.24 &     12.189 &      0.023 &     11.563 &      0.022 &     11.429 &      0.021 &          g &          2 \\
    SA 127 &   11 03 37 &  -17 43 23 &13.81 &     11.840 &      0.021 &     11.278 &      0.024 &     11.134 &      0.023 &          g &          2 \\
    SA 127 &   11 04 25 &  -17 57 56 &15.65 &     13.310 &      0.026 &     12.605 &      0.029 &     12.420 &      0.023 &          g &          2 \\
    SA 127 &   11 04 42 &  -17 50 02 &14.01 &     11.921 &      0.024 &     11.246 &      0.023 &     11.108 &      0.023 &          g &          2 \\
    SA 127 &   11 04 55 &  -17 46 14 &15.15 &     13.175 &      0.024 &     12.493 &      0.027 &     12.488 &      0.026 &          g &          2 \\
    SA 127 &   11 04 57 &  -18 27 56 &15.02 &     13.238 &      0.023 &     12.708 &      0.030 &     12.665 &      0.029 &          g &          2 \\
    SA 127 &   11 04 58 &  -18 07 19 &14.96 &     12.902 &      0.023 &     12.266 &      0.024 &     12.128 &      0.026 &          g &          2 \\
    SA 127 &   11 05 28 &  -18 54 55 &14.14 &     12.189 &      0.021 &     11.608 &      0.023 &     11.493 &      0.025 &          g &          2 \\
    SA 127 &   11 05 50 &  -16 53 02 &15.46 &     13.738 &      0.026 &     13.249 &      0.032 &     13.110 &      0.040 &          g &          2 \\
    SA 127 &   11 05 53 &  -19 34 51 &14.30 &     12.009 &      0.021 &     11.333 &      0.025 &     11.229 &      0.023 &          g &          2 \\
    SA 127 &   11 06 28 &  -19 47 00 &14.92 &     12.827 &      0.022 &     12.212 &      0.027 &     12.089 &      0.026 &          g &          2 \\
    SA 127 &   11 07 03 &  -18 06 17 &15.56 &     13.784 &      0.029 &     13.229 &      0.035 &     13.209 &      0.037 &          g &          2 \\
    SA 127 &   11 07 39 &  -19 28 31 &15.39 &     13.288 &      0.025 &     12.662 &      0.029 &     12.463 &      0.030 &          g &          2 \\
 HD 096360 &   11 07 53 &  +68 21 59 & 8.10 &      4.069 &      0.192 &      3.166 &      0.190 &      2.768 &      0.250 &          g &          1 \\
    SA 127 &   11 08 17 &  -16 31 35 &14.47 &     12.648 &      0.036 &     11.897 &      0.027 &     11.789 &      0.024 &          g &          2 \\
    SA 127 &   11 08 21 &  -17 53 37 &15.22 &     13.494 &      0.026 &     13.011 &      0.028 &     12.846 &      0.033 &          g &          2 \\
    SA 127 &   11 08 30 &  -19 56 10 &14.33 &     12.063 &      0.022 &     11.382 &      0.022 &     11.261 &      0.025 &          g &          2 \\
    SA 127 &   11 08 39 &  -17 20 07 &14.82 &     12.262 &      0.023 &     11.552 &      0.023 &     11.397 &      0.021 &          g &          2 \\
    SA 127 &   11 09 14 &  -19 35 29 &13.77 &     11.902 &      0.023 &     11.346 &      0.025 &     11.239 &      0.026 &          g &          2 \\
    SA 127 &   11 09 43 &  -19 27 26 &15.06 &     12.701 &      0.023 &     11.947 &      0.025 &     11.848 &      0.025 &          g &          2 \\
    SA 127 &   11 09 53 &  -18 34 10 &15.35 &     13.475 &      0.025 &     12.963 &      0.030 &     12.863 &      0.030 &          g &          2 \\
    SA 127 &   11 10 24 &  -18 50 29 &14.68 &     12.837 &      0.023 &     12.246 &      0.023 &     12.148 &      0.026 &          g &          2 \\
    SA 127 &   11 11 19 &  -17 03 52 &15.40 &     13.713 &      0.023 &     13.179 &      0.029 &     13.157 &      0.035 &          g &          2 \\
    SA 127 &   11 11 25 &  -19 15 25 &14.52 &     12.518 &      0.026 &     11.869 &      0.022 &     11.829 &      0.021 &          g &          2 \\
    SA 127 &   11 11 34 &  -16 15 18 &13.77 &     11.650 &      0.023 &     11.004 &      0.027 &     10.899 &      0.023 &          g &          2 \\
    SA 127 &   11 11 59 &  -19 07 36 &13.85 &     12.094 &      0.023 &     11.564 &      0.022 &     11.523 &      0.023 &          g &          2 \\
    SA 127 &   11 12 02 &  -19 21 01 &14.92 &     12.785 &      0.022 &     12.155 &      0.023 &     12.024 &      0.024 &          g &          2 \\
    SA 127 &   11 12 03 &  -18 54 32 &15.27 &     13.403 &      0.027 &     12.805 &      0.025 &     12.694 &      0.029 &          g &          2 \\
    SA 127 &   11 12 16 &  -16 11 13 &15.45 &     13.711 &      0.028 &     13.145 &      0.029 &     13.066 &      0.039 &          g &          2 \\
    SA 127 &   11 12 17 &  -16 08 40 &13.56 &     11.639 &      0.024 &     11.098 &      0.023 &     11.018 &      0.021 &          g &          2 \\
BD +36 2165 &  11 12 48 &  +35 43 44 & 9.75 &      8.791 &      0.027 &      8.552 &      0.031 &      8.498 &      0.018 &          d &          1 \\
LP 0732-0048&  11 13 05 &  -12 47 54 &12.51 &     11.532 &      0.024 &     11.264 &      0.026 &     11.242 &      0.021 &          d &          1 \\
    SA 127 &   11 13 18 &  -17 39 34 &15.67 &     13.655 &      0.026 &     13.042 &      0.022 &     12.860 &      0.030 &          g &          2 \\
    SA 127 &   11 13 41 &  -16 44 32 &13.96 &     11.584 &      0.022 &     10.944 &      0.023 &     10.796 &      0.026 &          g &          2 \\
    SA 127 &   11 14 06 &  -16 34 24 &13.69 &     11.588 &      0.022 &     10.986 &      0.022 &     10.867 &      0.021 &          g &          2 \\
    SA 127 &   11 14 17 &  -16 43 28 &15.17 &     13.112 &      0.024 &     12.482 &      0.023 &     12.392 &      0.027 &          g &          2 \\
\end{tabular}  
}  
\end{table*}

\begin{table*}
{\tiny
\begin{tabular}{lcccrrrrrrcr}
\hline
Star/field &  $\alpha$ & $\delta$ &     V & \multicolumn{2} {c} {J} & \multicolumn{2} {c} {H} & \multicolumn{2} {c} {K}&       type &       refs \\
&  (h~~m~~s) & ($^{o}~^{'}~^{''}$) &     (mag) & \multicolumn{2} {c} {(mag)} & \multicolumn{2} {c} {(mag)} & \multicolumn{2} {c} {(mag)}&    &     \\
\hline
    SA 127 &   11 14 31 &  -19 17 33 &15.55 &     13.440 &$\pm$ 0.026 &     12.800 &$\pm$ 0.026 &     12.673 & $\pm$0.027 &          g &          2 \\
    SA 127 &   11 15 48 &  -17 29 11 &15.33 &     13.470 &      0.027 &     12.952 &      0.025 &     12.844 &      0.029 &          g &          2 \\
 HD 097916 &   11 15 54 &  +02 05 12 & 9.17 &      8.319 &      0.018 &      8.104 &      0.031 &      8.018 &      0.029 &          d &          1 \\
    SA 127 &   11 16 23 &  -19 12 34 &14.81 &     12.457 &      0.024 &     11.708 &      0.022 &     11.614 &      0.025 &          g &          2 \\
    SA 127 &   11 16 26 &  -17 50 38 &15.51 &     13.729 &      0.028 &     13.157 &      0.023 &     13.066 &      0.031 &          g &          2 \\
    SA 127 &   11 16 56 &  -17 20 07 &15.30 &     13.520 &      0.029 &     12.966 &      0.031 &     12.935 &      0.038 &          g &          2 \\
    SA 127 &   11 17 02 &  -16 48 45 &15.79 &     13.663 &      0.024 &     12.983 &      0.025 &     12.799 &      0.028 &          g &          2 \\
    SA 127 &   11 17 42 &  -19 57 37 &15.60 &     13.327 &      0.024 &     12.629 &      0.029 &     12.447 &      0.023 &          g &          2 \\
    SA 127 &   11 18 15 &  -19 08 13 &14.52 &     12.764 &      0.030 &     12.200 &      0.024 &     12.122 &      0.026 &          g &          2 \\
    SA 127 &   11 18 38 &  -16 37 27 &14.19 &     12.192 &      0.025 &     11.573 &      0.023 &     11.459 &      0.022 &          g &          2 \\
    SA 127 &   11 19 24 &  -16 45 52 &14.55 &     12.412 &      0.020 &     11.740 &      0.024 &     11.632 &      0.022 &          g &          2 \\
    SA 127 &   11 19 32 &  -18 01 55 &14.36 &     11.730 &      0.028 &     11.060 &      0.024 &     10.870 &      0.019 &          g &          2 \\
    SA 127 &   11 19 44 &  -18 59 53 &14.03 &     11.906 &      0.027 &     11.229 &      0.024 &     11.106 &      0.025 &          g &          2 \\
    SA 127 &   11 19 54 &  -17 33 37 &14.78 &     12.617 &      0.023 &     11.995 &      0.025 &     11.838 &      0.026 &          g &          2 \\
    SA 127 &   11 20 09 &  -19 23 33 &15.64 &     13.672 &      0.023 &     13.182 &      0.030 &     13.021 &      0.027 &          g &          2 \\
 HD 099109 &   11 24 17 &  -01 31 45 & 8.80 &      7.626 &      0.026 &      7.259 &      0.040 &      7.162 &      0.024 &          d &          1 \\
 HD 099383 &   11 25 50 &  -38 52 17 & 9.08 &      8.025 &      0.023 &      7.756 &      0.044 &      7.644 &      0.033 &          d &          1 \\
BD +21 2321 &   11 27 24 &  +20 42 06&11.52 &     10.188 &      0.022 &      9.836 &      0.023 &      9.766 &      0.017 &          d &          1 \\
 HD 100503 &   11 33 47 &  -31 05 17 & 8.73 &      6.151 &      0.023 &      5.391 &      0.029 &      5.174 &      0.024 &          g &          1 \\
BD -02 3375 &   11 35 42 &  -03 39 28&11.00 &      9.871 &      0.023 &      9.593 &      0.024 &      9.518 &      0.023 &          d &          1 \\
 HD 100906 &   11 36 42 &  -18 58 11 & 9.66 &      8.031 &      0.019 &      7.595 &      0.021 &      7.493 &      0.020 &          g &          1 \\
 HD 101227 &   11 39 06 &  +44 18 20 & 8.39 &      7.128 &      0.023 &      6.807 &      0.038 &      6.736 &      0.020 &          d &          1 \\
BD +26 2251 &   11 44 36 &  +25 32 12&10.35 &      9.290 &      0.018 &      9.022 &      0.016 &      8.963 &      0.018 &          d &          1 \\
 HD 102158 &   11 45 31 &  +47 40 01 & 8.03 &      6.860 &      0.026 &      6.589 &      0.020 &      6.509 &      0.026 &          d &          1 \\
BD +51 1696 &   11 46 35 &  +50 52 55& 9.90 &      8.678 &      0.020 &      8.375 &      0.027 &      8.306 &      0.020 &          d &          1 \\
BD -13 3442 &   11 46 51 &  -14 06 43&10.37 &      9.293 &      0.026 &      9.034 &      0.026 &      9.018 &      0.019 &          d &          1 \\
 HD 103036 &   11 51 50 &  -05 45 44 & 8.18 &      5.860 &      0.020 &      5.247 &      0.029 &      5.046 &      0.021 &          g &          1 \\
 HD 103295 &   11 53 37 &  -28 38 13 & 9.56 &      7.945 &      0.026 &      7.498 &      0.034 &      7.374 &      0.018 &          g &          1 \\
BD -01 2582 &   11 53 37 &  -02 00 37& 9.62 &      8.126 &      0.020 &      7.717 &      0.067 &      7.598 &      0.021 &          g &          1 \\
 HD 103545 &   11 55 27 &  +09 07 45 & 9.46 &      7.608 &      0.021 &      7.104 &      0.034 &      6.988 &      0.029 &          g &          1 \\
BD -21 3420 &   11 55 28 &  -22 23 13&10.15 &      9.042 &      0.020 &      8.703 &      0.042 &      8.698 &      0.023 &          d &          1 \\
 HD 103723 &   11 56 36 &  -21 25 10 &10.07 &      9.007 &      0.027 &      8.718 &      0.040 &      8.655 &      0.021 &          d &          1 \\
 HD 103912 &   11 58 00 &  +48 12 12 & 8.39 &      6.681 &      0.019 &      6.261 &      0.024 &      6.159 &      0.018 &          g &          1 \\
 HD 233891 &   11 59 59 &  +51 46 18 & 8.84 &      7.159 &      0.018 &      6.694 &      0.020 &      6.607 &      0.017 &          g &          1 \\
 HD 104340 &   12 00 57 &  -21 15 03 & 8.17 &      5.853 &      0.019 &      5.219 &      0.021 &      5.037 &      0.020 &          g &          1 \\
 HD 104893 &   12 04 43 &  -29 11 05 & 9.25 &      7.001 &      0.018 &      6.384 &      0.033 &      6.257 &      0.020 &          g &          1 \\
 HD 105004 &   12 05 25 &  -26 35 44 &10.21 &      9.229 &      0.029 &      8.949 &      0.051 &      8.865 &      0.023 &          d &          1 \\
 HD 105546 &   12 09 03 &  +59 01 05 & 8.64 &      7.152 &      0.026 &      6.756 &      0.024 &      6.674 &      0.018 &          g &          1 \\
 HD 105755 &   12 10 16 &  +54 29 17 & 8.59 &      7.414 &      0.020 &      7.126 &      0.024 &      7.062 &      0.024 &          d &          1 \\
 HD 105740 &   12 10 17 &  +16 22 13 & 8.38 &      6.506 &      0.018 &      5.963 &      0.020 &      5.847 &      0.024 &          g &          1 \\
 HD 106038 &   12 12 01 &  +13 15 41 &10.18 &      9.107 &      0.029 &      8.834 &      0.061 &      8.761 &      0.018 &          d &          1 \\
 HD 106191 &   12 13 11 &  -15 13 56 &10.00 &      8.935 &      0.022 &      8.672 &      0.023 &      8.586 &      0.022 &          d &          1 \\
 HD 107752 &   12 22 53 &  +11 36 25 &10.07 &      8.252 &      0.026 &      7.754 &      0.040 &      7.661 &      0.016 &          g &          1 \\
BD +31 2360&   12 24 18 &  +30 46 00 &11.28 &      9.449 &      0.018 &      8.894 &      0.015 &      8.830 &      0.016 &          g &          1 \\
 HD 108076 &   12 24 46 &  +38 19 07 & 8.02 &      6.815 &      0.019 &      6.542 &      0.020 &      6.483 &      0.020 &          d &          1 \\
 HD 108177 &   12 25 35 &  +01 17 02 & 9.66 &      8.673 &      0.023 &      8.404 &      0.033 &      8.354 &      0.026 &          d &          1 \\
 HD 108317 &   12 26 37 &  +05 18 09 & 8.04 &      6.623 &      0.026 &      6.233 &      0.031 &      6.153 &      0.023 &          g &          1 \\
 HD 108577 &   12 28 17 &  +12 20 41 & 9.55 &      7.995 &      0.024 &      7.549 &      0.016 &      7.463 &      0.031 &          g &          1 \\
BD +04 2621 &  12 28 45 &  +04 01 27 & 9.91 &      8.179 &      0.023 &      7.688 &      0.040 &      7.549 &      0.020 &          g &          1 \\
 HD 108754 &   12 29 43 &  -03 19 59 & 9.03 &      7.660 &      0.023 &      7.295 &      0.027 &      7.201 &      0.026 &          d &          1 \\
CD -39 07674&  12 32 29 &  -40 05 55 &11.10 &      9.893 &      0.026 &      9.659 &      0.025 &      9.613 &      0.023 &          d &          1 \\
 HD 109303 &   12 33 19 &  +49 18 07 & 8.15 &      7.045 &      0.021 &      6.822 &      0.031 &      6.757 &      0.017 &          d &          1 \\
 HD 109443 &   12 34 47 &  -23 28 32 & 9.25 &      8.404 &      0.024 &      8.210 &      0.051 &      8.157 &      0.026 &          d &          1 \\
 HD 110184 &   12 40 14 &  +08 31 38 & 8.31 &      6.125 &      0.024 &      5.505 &      0.047 &      5.346 &      0.026 &          g &          1 \\
 HD 110281 &   12 41 04 &  +00 37 14 & 9.39 &      6.653 &      0.023 &      5.910 &      0.027 &      5.775 &      0.018 &          g &          1 \\
MFF90 PHI 4-34&12 43 15 &  -28 56 25 &11.63 &      9.850 &      0.022 &      9.371 &      0.025 &      9.242 &      0.022 &          g &          1 \\
BD +03 2688 &  12 44 08 &  +02 44 38 &10.50 &      8.520 &      0.029 &      7.976 &      0.042 &      7.911 &      0.029 &          g &          1 \\
 HD 111515 &   12 49 45 &  +01 11 17 & 8.10 &      6.812 &      0.023 &      6.493 &      0.053 &      6.358 &      0.020 &          d &          1 \\
MFF90 PHI 4-17&12 52 17 &  -30 01 54 &11.91 &      9.573 &      0.024 &      8.875 &      0.025 &      8.715 &      0.023 &          g &          1 \\
 HD 111980 &   12 53 15 &  -18 31 20 & 8.38 &      7.176 &      0.024 &      6.890 &      0.047 &      6.768 &      0.020 &          d &          1 \\
 HD 113083 &   13 01 26 &  -27 22 28 & 8.05 &      6.964 &      0.023 &      6.659 &      0.033 &      6.576 &      0.018 &          d &          1 \\
 HD 113679 &   13 05 53 &  -38 30 60 & 9.70 &      8.466 &      0.018 &      8.228 &      0.046 &      8.108 &      0.029 &          d &          1 \\
 HD 113801 &   13 06 25 &  -20 03 31 & 8.45 &      6.631 &      0.019 &      6.160 &      0.031 &      5.985 &      0.018 &          g &          1 \\
BD +33 2300 &  13 06 33 &  +32 40 01 &10.09 &      9.044 &      0.022 &      8.803 &      0.019 &      8.725 &      0.017 &          d &          1 \\
 HD 114095 &   13 08 26 &  -07 18 30 & 8.35 &      6.541 &      0.018 &      6.043 &      0.040 &      5.887 &      0.020 &          g &          1 \\
 HD 114606 &   13 11 21 &  +09 37 34 & 8.74 &      7.534 &      0.019 &      7.219 &      0.036 &      7.088 &      0.024 &          d &          1 \\
 HD 115444 &   13 16 42 &  +36 22 53 & 9.00 &      7.160 &      0.019 &      6.702 &      0.023 &      6.607 &      0.020 &          g &          1 \\
 HD 115772 &   13 20 00 &  -39 56 20 & 9.63 &      8.090 &      0.041 &      7.622 &      0.049 &      7.589 &      0.020 &          g &          1 \\
 HD 116064 &   13 21 44 &  -39 18 40 & 8.81 &      7.698 &      0.024 &      7.372 &      0.038 &      7.306 &      0.024 &          d &          1 \\
 G 255-032 &   13 21 48 &  +74 12 33 &11.64 &     10.439 &      0.021 &     10.118 &      0.021 &     10.070 &      0.022 &          d &          1 \\
 HD 117243 &   13 28 39 &  +28 26 55 & 8.34 &      7.200 &      0.023 &      6.930 &      0.021 &      6.879 &      0.024 &          d &          1 \\
 HD 117220 &   13 29 36 &  -37 33 36 & 9.04 &      7.386 &      0.018 &      6.874 &      0.036 &      6.802 &      0.020 &          g &          1 \\
BD +03 2782 &  13 29 56 &  +02 45 27 & 9.72 &      7.734 &      0.029 &      7.187 &      0.053 &      7.077 &      0.034 &          g &          1 \\
 HD 118055 &   13 34 40 &  -16 19 23 & 8.89 &      6.573 &      0.020 &      5.911 &      0.049 &      5.724 &      0.020 &          g &          1 \\
BD +01 2831 &  13 36 02 &  +01 12 08 &10.89 &      9.455 &      0.026 &      9.063 &      0.026 &      8.993 &      0.023 &          d &          1 \\
BD +18 2757 &  13 36 32 &  +18 09 02 & 9.83 &      8.094 &      0.019 &      7.679 &      0.017 &      7.619 &      0.021 &          g &          1 \\
 HD 118659 &   13 38 00 &  +19 08 53 & 8.84 &      7.553 &      0.039 &      7.194 &      0.024 &      7.132 &      0.021 &          d &          1 \\
 G 064-012 &   13 40 02 &  +00 02 19 &11.49 &     10.509 &      0.024 &     10.268 &      0.023 &     10.208 &      0.021 &          d &          1 \\
 HD 119173 &   13 41 43 &  -04 01 46 & 8.80 &      7.730 &      0.024 &      7.460 &      0.027 &      7.390 &      0.024 &          d &          1 \\
 HD 119667 &   13 44 45 &  -03 32 00 & 8.45 &      5.269 &      0.021 &      4.408 &      0.210 &      4.185 &      0.238 &          g &          1 \\
 HD 120170 &   13 47 59 &  -08 47 23 & 9.03 &      7.257 &      0.026 &      6.751 &      0.038 &      6.613 &      0.018 &          g &          1 \\
 HD 121135 &   13 53 34 &  +02 41 41 & 9.41 &      7.750 &      0.018 &      7.300 &      0.034 &      7.182 &      0.023 &          g &          1 \\
 HD 121258 &   13 54 55 &  -26 00 57 &10.50 &      9.272 &      0.024 &      8.885 &      0.042 &      8.803 &      0.021 &          d &          1 \\
BD +34 2476 &  13 59 09 &  +33 51 39 &10.06 &      9.077 &      0.019 &      8.850 &      0.016 &      8.811 &      0.018 &          d &          1 \\
 HD 122196 &   14 01 02 &  -38 03 03 & 8.75 &      7.629 &      0.027 &      7.361 &      0.033 &      7.275 &      0.024 &          d &          1 \\
 HD 122547 &   14 01 55 &  +32 49 33 & 9.42 &      7.361 &      0.021 &      6.790 &      0.021 &      6.605 &      0.018 &          g &          1 \\
 G 064-037 &   14 02 30 &  -05 39 05 &11.14 &     10.188 &      0.028 &      9.956 &      0.024 &      9.923 &      0.023 &          d &          1 \\
 HD 123710 &   14 04 57 &  +74 34 25 & 8.22 &      6.999 &      0.020 &      6.703 &      0.026 &      6.650 &      0.020 &          d &          1 \\
 HD 123821 &   14 08 27 &  +51 35 33 & 8.63 &      6.952 &      0.020 &      6.540 &      0.026 &      6.379 &      0.016 &          g &          1 \\
 HD 124244 &   14 12 17 &  +08 24 25 & 8.48 &      7.293 &      0.029 &      7.029 &      0.047 &      6.908 &      0.024 &          d &          1 \\
BD +09 2870 &  14 16 30 &  +08 27 53 & 9.45 &      7.534 &      0.020 &      6.968 &      0.016 &      6.869 &      0.023 &          g &          1 \\
 G 239-012 &   14 18 53 &  +73 14 12 &11.61 &     10.628 &      0.022 &     10.361 &      0.028 &     10.291 &      0.020 &          d &          1 \\
BD +01 2916 &  14 21 45 &  +00 46 59 & 9.65 &      7.273 &      0.030 &      6.566 &      0.023 &      6.472 &      0.026 &          g &          1 \\
BD +08 2856 &  14 23 58 &  +08 01 33 & 9.96 &      8.060 &      0.020 &      7.530 &      0.051 &      7.387 &      0.033 &          g &          1 \\
 HD 126511 &   14 24 49 &  +41 16 30 & 8.30 &      7.001 &      0.024 &      6.677 &      0.018 &      6.606 &      0.020 &          d &          1 \\
 HD 126614 &   14 26 48 &  -05 10 40 & 8.70 &      7.470 &      0.021 &      7.160 &      0.042 &      7.060 &      0.036 &          d &          1 \\
 HD 126587 &   14 27 00 &  -22 14 39 & 9.15 &      7.258 &      0.020 &      6.777 &      0.042 &      6.668 &      0.020 &          g &          1 \\
 HD 126681 &   14 27 25 &  -18 24 40 & 9.32 &      8.044 &      0.023 &      7.709 &      0.040 &      7.631 &      0.024 &          d &          1 \\
BD +18 2890 &  14 32 13 &  +17 25 24 & 9.77 &      8.241 &      0.023 &      7.837 &      0.018 &      7.744 &      0.018 &          g &          1 \\
 G 166-037 &   14 34 51 &  +25 09 54 &12.66 &     11.246 &      0.022 &     10.839 &      0.023 &     10.775 &      0.014 &          d &          1 \\
 G 066-009 &   14 35 14 &  +12 13 30 &12.01 &     10.827 &      0.021 &     10.563 &      0.019 &     10.489 &      0.018 &          d &          1 \\
 HD 128188 &   14 35 47 &  -11 24 12 & 9.94 &      8.089 &      0.024 &      7.516 &      0.018 &      7.384 &      0.024 &          g &          1 \\
 HD 129392 &   14 42 11 &  +03 56 19 & 8.92 &      8.032 &      0.030 &      7.827 &      0.042 &      7.789 &      0.031 &          d &          1 \\
 HD 129515 &   14 42 11 &  +31 55 36 & 8.78 &      7.790 &      0.023 &      7.583 &      0.036 &      7.519 &      0.020 &          d &          1 \\
 HD 129518 &   14 42 49 &  +04 02 45 & 8.86 &      7.866 &      0.024 &      7.623 &      0.040 &      7.540 &      0.024 &          d &          1 \\
 HD 130322 &   14 47 33 &  +00 16 53 & 8.05 &      6.712 &      0.023 &      6.315 &      0.027 &      6.234 &      0.023 &          d &          1 \\
BD +26 2606 &  14 49 02 &  +25 42 09 & 9.72 &      8.676 &      0.027 &      8.394 &      0.023 &      8.352 &      0.020 &          d &          1 \\
 G 066-030 &   14 50 08 &  +00 50 27 &11.07 &     10.062 &      0.023 &      9.842 &      0.025 &      9.789 &      0.022 &          d &          1 \\
 HD 131653 &   14 55 07 &  -09 05 50 & 9.50 &      8.154 &      0.035 &      7.802 &      0.051 &      7.680 &      0.033 &          d &          1 \\
 HD 132475 &   14 59 50 &  -22 00 46 & 8.57 &      7.327 &      0.026 &      6.996 &      0.036 &      6.912 &      0.021 &          d &          1 \\
BD -11 3853 &  15 00 31 &  -12 26 57 &10.21 &      9.089 &      0.025 &      8.797 &      0.025 &      8.733 &      0.028 &          d &          1 \\
BPS BS 16968-0061&15 01 06&+03 42 46 &13.26 &     12.214 &      0.024 &     11.929 &      0.022 &     11.866 &      0.025 &          d &          1 \\
BD +30 2611 &  15 06 54 &  +30 00 37 & 9.14 &      6.882 &      0.023 &      6.203 &      0.018 &      6.094 &      0.021 &          g &          1 \\
 HD 134113 &   15 07 46 &  +08 52 47 & 8.26 &      7.093 &      0.030 &      6.761 &      0.031 &      6.695 &      0.017 &          d &          1 \\
 HD 134088 &   15 08 13 &  -07 54 48 & 8.00 &      6.807 &      0.029 &      6.490 &      0.047 &      6.434 &      0.033 &          d &          1 \\
\end{tabular}  
}  
\end{table*}

\begin{table*}
{\tiny
\begin{tabular}{lcccrrrrrrcr}
\hline
Star/field &  $\alpha$ & $\delta$ &     V & \multicolumn{2} {c} {J} & \multicolumn{2} {c} {H} & \multicolumn{2} {c} {K}&       type &       refs \\
&  (h~~m~~s) & ($^{o}~^{'}~^{''}$) &     (mag) & \multicolumn{2} {c} {(mag)} & \multicolumn{2} {c} {(mag)} & \multicolumn{2} {c} {(mag)}&    &     \\
\hline
 HD 135148 &   15 13 17 &  +12 27 26 & 9.38 &      7.137 &$\pm$ 0.030 &      6.515 &$\pm$ 0.020 &      6.368 &$\pm$ 0.023 &          g &          1 \\
 G 152-035 &   15 32 00 &  -11 22 42 &12.15 &     10.814 &      0.022 &     10.493 &      0.022 &     10.419 &      0.019 &          d &          1 \\
 HD 138776 &   15 34 17 &  -02 43 27 & 8.72 &      7.475 &      0.018 &      7.198 &      0.033 &      7.112 &      0.027 &          d &          1 \\
 HD 141531 &   15 49 16 &  +09 36 42 & 9.15 &      6.971 &      0.027 &      6.337 &      0.034 &      6.215 &      0.020 &          g &          1 \\
BPS CS 22884-0108&15 49 57&-09 14 11 &14.24 &     12.996 &      0.027 &     12.679 &      0.023 &     12.588 &      0.026 &          d &          1 \\
 G 016-025 &   16 01 23 &  +05 24 00 &13.33 &     12.051 &      0.024 &     11.674 &      0.025 &     11.626 &      0.027 &          d &          1 \\
BD +42 2667 &  16 03 13 &  +42 14 47 & 9.85 &      8.822 &      0.023 &      8.517 &      0.021 &      8.491 &      0.022 &          d &          1 \\
 HD 144921 &   16 07 27 &  +24 54 30 & 8.35 &      5.538 &      0.052 &      5.023 &      0.166 &      4.638 &      0.016 &          g &          1 \\
 HD 146099 &   16 12 43 &  +43 38 59 & 8.13 &      7.095 &      0.029 &      6.827 &      0.017 &      6.765 &      0.021 &          d &          1 \\
 HD 147609 &   16 21 52 &  +27 22 27 & 9.21 &      8.211 &      0.024 &      8.035 &      0.036 &      7.948 &      0.020 &          d &          1 \\
 HD 148408 &   16 27 48 &  -01 04 09 & 9.62 &      8.252 &      0.023 &      7.876 &      0.053 &      7.781 &      0.026 &          d &          1 \\
 G 180-058 &   16 28 17 &  +44 40 38 &11.29 &      9.901 &      0.022 &      9.507 &      0.021 &      9.397 &      0.018 &          d &          1 \\
BD +11 2998 &  16 30 17 &  +10 59 52 & 9.07 &      7.619 &      0.019 &      7.271 &      0.027 &      7.185 &      0.021 &          g &          1 \\
BD +09 3223 &  16 33 36 &  +09 06 16 & 9.25 &      7.760 &      0.020 &      7.335 &      0.040 &      7.277 &      0.023 &          g &          1 \\
 HD 149750 &   16 35 15 &  +37 01 20 & 8.59 &      7.416 &      0.023 &      7.132 &      0.027 &      7.044 &      0.023 &          d &          1 \\
 G 169-021 &   16 37 05 &  +31 19 24 & 12.12&     10.786 &      0.020 &     10.440 &      0.022 &     10.359 &      0.019 &          d &          1 \\
 HD 149996 &   16 38 17 &  -02 26 32 &  8.49&      7.325 &      0.020 &      7.028 &      0.044 &      6.922 &      0.023 &          d &          1 \\
LP 0685-0047 & 16 43 10 &  -05 56 48 & 12.55&     10.686 &      0.023 &     10.274 &      0.023 &     10.149 &      0.022 &          g &          1 \\
BPS CS 22878-0101&16 45 31&+08 14 46 & 13.79&     11.873 &      0.019 &     11.369 &      0.027 &     11.242 &      0.021 &          g &          1 \\
 G 139-008 &   17 01 44 &  +16 09 03 & 11.46&     10.333 &      0.018 &     10.066 &      0.017 &     10.007 &      0.017 &          d &          1 \\
 HD 158226 &   17 26 43 &  +31 04 38 &  8.50&      7.318 &      0.020 &      7.007 &      0.020 &      6.930 &      0.038 &          d &          1 \\
BD +17 3248 &  17 28 14 &  +17 30 36 &  9.37&      7.876 &      0.056 &      7.391 &      0.038 &      7.338 &      0.018 &          g &          1 \\
BD +23 3130 &  17 32 42 &  +23 44 12 &  8.95&      7.420 &      0.019 &      7.026 &      0.017 &      6.952 &      0.018 &          g &          1 \\
 HD 160693 &   17 39 37 &  +37 11 02 &  8.36&      7.226 &      0.021 &      6.957 &      0.026 &      6.896 &      0.016 &          d &          1 \\
BD +20 3603 &  17 54 43 &  +20 16 16 &  9.69&      8.860 &      0.020 &      8.620 &      0.016 &      8.603 &      0.020 &          d &          1 \\
BPS CS 22959-0007&18 41 13&-67 15 28 & 14.07&     13.213 &      0.026 &     13.011 &      0.025 &     12.952 &      0.033 &          d &          1 \\
 HD 178443 &   19 10 37 &  -43 16 36 & 10.04&      8.414 &      0.018 &      8.013 &      0.020 &      7.911 &      0.026 &          g &          1 \\
 HD 181743 &   19 23 43 &  -45 04 57 &  9.71&      8.624 &      0.023 &      8.348 &      0.020 &      8.274 &      0.021 &          d &          1 \\
 HD 187216 &   19 24 18 &  +85 21 57 &  9.56&      6.891 &      0.023 &      6.251 &      0.040 &      6.025 &      0.020 &          g &          1 \\
BPS CS 22891-0200 &19 35 19&-61 42 25& 13.93&     11.995 &      0.022 &     11.461 &      0.025 &     11.414 &      0.023 &          g &          1 \\
 HD 184711 &   19 37 12 &  -39 44 37 &  8.05&      5.513 &      0.024 &      4.862 &      0.076 &      4.708 &      0.016 &          g &          1 \\
BPS CS 22891-0209&19 42 02&-61 03 46 & 12.17&     10.235 &      0.024 &      9.745 &      0.022 &      9.623 &      0.021 &          g &          1 \\
BPS CS 22896-0154 &19 42 27&-56 58 34& 13.64&     12.072 &      0.026 &     11.637 &      0.023 &     11.540 &      0.024 &          g &          1 \\
BPS CS 22873-0055 &19 53 50&-59 40 00& 12.65&     10.654 &      0.024 &     10.127 &      0.022 &     10.006 &      0.022 &          g &          1 \\
BD -18 5550 &   19 58 50 &  -18 12 11&  9.35&      7.202 &      0.020 &      6.721 &      0.047 &      6.555 &      0.018 &          g &          1 \\
 HD 188985 &   19 59 58 &  -48 58 31 &  8.57&      7.546 &      0.021 &      7.320 &      0.031 &      7.231 &      0.027 &          d &          1 \\
 HD 190287 &   20 05 38 &  -34 55 12 &  8.54&      6.970 &      0.032 &      6.519 &      0.031 &      6.441 &      0.017 &          g &          1 \\
BPS CS 22873-0128&20 07 04&-58 34 57 & 13.03&     11.430 &      0.029 &     10.993 &      0.026 &     10.874 &      0.024 &          g &          1 \\
 HD 192031 &   20 13 23 &  -15 25 56 &  8.66&      7.305 &      0.020 &      6.873 &      0.024 &      6.813 &      0.020 &          d &          1 \\
BPS CS 22873-0166&20 19 22&-61 30 20 & 11.82&      9.832 &      0.024 &      9.263 &      0.026 &      9.137 &      0.023 &          g &          1 \\
BPS CS 22885-0096 &20 20 51&-39 53 30& 13.33&     11.630 &      0.024 &     11.163 &      0.025 &     11.090 &      0.025 &          g &          1 \\
BPS CS 22950-0046 &20 21 28&-13 16 36& 14.22&     12.250 &      0.022 &     11.685 &      0.022 &     11.583 &      0.026 &          g &          1 \\
 HD 193901 &   20 23 36 &  -21 22 14 &  8.67&      7.500 &      0.018 &      7.219 &      0.038 &      7.144 &      0.023 &          d &          1 \\
LP 0635-0014 & 20 26 47 &  +00 37 06 & 11.33&     10.285 &      0.023 &     10.041 &      0.022 &      9.997 &      0.020 &          d &          1 \\
 HD 195598 &   20 33 47 &  -39 36 23 &  9.24&      8.188 &      0.026 &      7.928 &      0.038 &      7.884 &      0.031 &          d &          1 \\
LP 0815-0043 & 20 38 14& -20 25 54   & 10.91&      9.964 &      0.022 &      9.709 &      0.022 &      9.650 &      0.020 &          d &          1 \\
 HD 196892 &   20 40 49 &  -18 47 33 &  8.23&      7.182 &      0.026 &      6.907 &      0.047 &      6.824 &      0.017 &          d &          1 \\
BD -15 5781 &  20 45 35 &  -14 31 15 & 10.80&      8.936 &      0.029 &      8.465 &      0.059 &      8.288 &      0.021 &          g &          1 \\
BD +04 4551 &  20 48 51 &  +05 11 59 &  9.69&      8.619 &      0.017 &      8.298 &      0.029 &      8.251 &      0.021 &          d &          1 \\
 HD 198245 &   20 50 23 &  -40 36 30 &  8.98&      7.749 &      0.018 &      7.459 &      0.053 &      7.321 &      0.027 &          d &          1 \\
BD -14 5890 &  20 56 09 &  -13 31 18 & 10.31&      8.511 &      0.024 &      8.022 &      0.023 &      7.939 &      0.036 &          g &          1 \\
 HD 199289 &   20 58 09 &  -48 12 13 &  8.30&      7.178 &      0.020 &      6.920 &      0.046 &      6.841 &      0.024 &          d &          1 \\
BPS CS 22897-0008&21 03 11&-65 05 15 & 13.33&     11.596 &      0.026 &     11.089 &      0.022 &     11.002 &      0.021 &          g &          1 \\
 HD 200654 &   21 06 35 &  -49 57 50 &  9.11&      7.648 &      0.018 &      7.252 &      0.036 &      7.154 &      0.020 &          g &          1 \\
 G 025-022 &   21 14 48 &  +11 00 30 & 12.32&     11.097 &      0.024 &     10.741 &      0.024 &     10.704 &      0.021 &          d &          1 \\
 HD 202206 &   21 14 58 &  -20 47 21 &  8.08&      6.850 &      0.026 &      6.567 &      0.016 &      6.485 &      0.023 &          d &          1 \\
 G 093-001 &   21 18 32 &  +02 36 24 & 14.64&     13.316 &      0.028 &     12.911 &      0.022 &     12.838 &      0.027 &          d &          1 \\
MFF90 959 11 094 &21 23 43&+06 22 07 & 11.24&      9.571 &      0.024 &      9.106 &      0.026 &      8.990 &      0.024 &          g &          1 \\
MFF90 959 11 058&21 23 59& +06 51 50 & 11.92&     10.270 &      0.023 &      9.801 &      0.027 &      9.681 &      0.025 &          g &          1 \\
MFF90 959 11 051 &21 25 09 &+06 56 31& 11.47&      9.648 &      0.023 &      9.117 &      0.026 &      8.976 &      0.021 &          g &          1 \\
MFF90 959 11 163 &21 25 43 &+05 29 49& 12.31&      9.864 &      0.024 &      9.151 &      0.026 &      8.988 &      0.023 &          g &          1 \\
 HD 204155 &   21 26 43 &  +05 26 30 &  8.47&      7.360 &      0.021 &      7.032 &      0.033 &      7.011 &      0.029 &          d &          1 \\
MFF90 959 11 003&21 26 46& +07 29 27 & 11.52&      9.795 &      0.026 &      9.218 &      0.024 &      9.149 &      0.026 &          g &          1 \\
MFF90 959 32 087 &21 28 32&+04 14 05 & 11.38&      9.054 &      0.027 &      8.317 &      0.057 &      8.193 &      0.017 &          g &          1 \\
 HD 204543 &   21 29 28 &  -03 30 55 &  8.31&      6.462 &      0.023 &      5.898 &      0.038 &      5.777 &      0.016 &          g &          1 \\
CD -35 14849 &   21 33 50 &-35 26 14 & 10.63&      9.590 &      0.023 &      9.355 &      0.025 &      9.293 &      0.022 &          d &          1 \\
 HD 205156 &   21 34 51 &  -49 47 35 &  8.16&      6.955 &      0.024 &      6.642 &      0.034 &      6.517 &      0.024 &          d &          1 \\
 HD 205650 &   21 37 26 &  -27 38 07 &  9.00&      7.917 &      0.020 &      7.628 &      0.027 &      7.573 &      0.023 &          d &          1 \\
BPS CS 22948-0027&21 37 46&-39 27 19 & 12.66&     10.979 &      0.024 &     10.534 &      0.021 &     10.427 &      0.023 &          g &          1 \\
 HD 206433 &   21 41 54 &  +10 04 23 &  9.39&      8.440 &      0.021 &      8.261 &      0.057 &      8.183 &      0.029 &          d &          1 \\
 HD 206739 &   21 44 24 &  -11 46 23 &  8.56&      6.698 &      0.021 &      6.138 &      0.029 &      6.033 &      0.024 &          g &          1 \\
BD -09 5831 &  21 45 31 &  -08 24 35 & 10.40&      8.486 &      0.020 &      7.926 &      0.023 &      7.813 &      0.026 &          g &          1 \\
 G 018-024 &   22 05 01 &  +08 38 18 & 12.03&     10.820 &      0.022 &     10.528 &      0.022 &     10.474 &      0.021 &          d &          1 \\
BD +11 4725&   22 05 41 &  +12 22 36 &  9.55&      8.223 &      0.021 &      7.856 &      0.018 &      7.757 &      0.020 &          d &          1 \\
 HD 210295 &   22 09 41 &  -13 36 19 &  9.59&      7.839 &      0.020 &      7.315 &      0.026 &      7.183 &      0.020 &          g &          1 \\
BD +17 4708 &  22 11 31 &  +18 05 34 &  9.47&      8.435 &      0.056 &      8.108 &      0.046 &      8.075 &      0.038 &          d &          1 \\
 HD 210631 &   22 11 39 &  +06 11 36 &  8.51&      7.269 &      0.020 &      6.944 &      0.049 &      6.891 &      0.027 &          d &          1 \\
 HD 211075 &   22 14 20 &  +18 01 13 &  8.21&      5.974 &      0.019 &      5.378 &      0.038 &      5.217 &      0.018 &          g &          1 \\
BPS CS 22892-0052&22 17 02&-16 39 26 & 13.18&     11.492 &      0.021 &     11.022 &      0.022 &     10.929 &      0.021 &          g &          1 \\
BD +07 4841 &  22 18 37 &  +08 26 45 & 10.38&      9.334 &      0.029 &      9.054 &      0.024 &      9.017 &      0.023 &          d &          1 \\
 HD 211744 &   22 20 10 &  -44 21 53 &  9.15&      7.642 &      0.024 &      7.206 &      0.046 &      7.117 &      0.024 &          g &          1 \\
BPS CS 29491-0084&22 28 50&-28 57 03 &13.48 &     12.524 &      0.023 &     12.251 &      0.026 &     12.201 &      0.028 &          d &          1 \\
 HD 213467 &   22 32 08 &  -31 10 25 & 8.58 &      7.041 &      0.027 &      6.549 &      0.027 &      6.479 &      0.027 &          g &          1 \\
 HD 213657 &   22 33 47 &  -42 03 14 & 9.66 &      8.672 &      0.026 &      8.406 &      0.027 &      8.346 &      0.026 &          d &          1 \\
 G 027-045 &   22 44 56 &  -02 21 13 &11.48 &     10.093 &      0.026 &      9.706 &      0.024 &      9.652 &      0.028 &          d &          1 \\
 HD 215601 &   22 46 48 &  -31 52 19 & 8.49 &      6.850 &      0.019 &      6.405 &      0.018 &      6.280 &      0.023 &          g &          1 \\
 HD 216179 &   22 50 46 &  +01 51 55 & 9.34 &      7.977 &      0.023 &      7.639 &      0.036 &      7.561 &      0.021 &          d &          1 \\
 HD 216777 &   22 55 50 &  -07 49 21 & 8.03 &      6.816 &      0.034 &      6.492 &      0.033 &      6.353 &      0.020 &          d &          1 \\
 HD 217272 &   23 00 16 &  -52 49 19 & 9.03 &      6.772 &      0.024 &      6.119 &      0.047 &      6.001 &      0.021 &          g &          1 \\
CD -24 17504 & 23 07 21 &  -23 52 36 &12.18 &     11.121 &      0.022 &     10.874 &      0.023 &     10.807 &      0.023 &          d &          1 \\
 HD 218502 &   23 08 39 &  -15 03 12 & 8.50 &      7.265 &      0.024 &      7.020 &      0.033 &      6.972 &      0.017 &          d &          1 \\
 HD 218504 &   23 08 43 &  -18 45 07 & 8.13 &      7.014 &      0.018 &      6.730 &      0.034 &      6.666 &      0.021 &          d &          1 \\
 HD 218857 &   23 11 25 &  -16 15 04 & 8.95 &      7.397 &      0.019 &      6.991 &      0.057 &      6.873 &      0.017 &          g &          1 \\
CPD -64 4333 & 23 16 31 &  -63 27 05 & 9.59 &      7.762 &      0.023 &      7.209 &      0.029 &      7.088 &      0.018 &          g &          1 \\
 HD 219617 &   23 17 05 &  -13 51 04 & 8.16 &      7.082 &      0.019 &      6.860 &      0.020 &      6.771 &      0.023 &          d &          1 \\
 HD 219715 &   23 18 01 &  +09 04 27 & 9.19 &      7.720 &      0.029 &      7.291 &      0.034 &      7.183 &      0.018 &          g &          1 \\
BD +02 4651 &  23 19 40 &  +03 22 17 &10.19 &      9.193 &      0.023 &      8.902 &      0.024 &      8.831 &      0.022 &          d &          1 \\
BD +33 4707 &  23 25 11 &  +34 17 14 & 9.00 &      7.628 &      0.020 &      7.153 &      0.024 &      7.062 &      0.017 &          d &          1 \\
 HD 220662 &   23 25 42 &  -23 56 21 &10.03 &      8.143 &      0.024 &      7.545 &      0.031 &      7.444 &      0.027 &          g &          1 \\
BPS CS 22949-0048&23 26 09&-05 49 58 &13.66 &     11.919 &      0.023 &     11.374 &      0.024 &     11.282 &      0.023 &          g &          1 \\
BPS CS 22949-0037&23 26 32&-02 39 47 &14.36 &     12.650 &      0.024 &     12.159 &      0.023 &     12.075 &      0.024 &          g &          1 \\
 HD 220838 &   23 27 16 &  -26 58 58 & 9.39 &      7.226 &      0.027 &      6.667 &      0.044 &      6.483 &      0.026 &          g &          1 \\
BD -08 6117 &   23 28 56 & -07 49 19 & 9.60 &      9.226 &      0.024 &      8.662 &      0.055 &      8.543 &      0.021 &          d &          1 \\
 HD 221580 &   23 33 40 &  -53 39 29 & 9.22 &      7.736 &      0.024 &      7.370 &      0.051 &      7.247 &      0.017 &          g &          1 \\
BPS CS 22952-0015 &23 37 29&-05 47 56&13.28 &     11.520 &      0.022 &     10.999 &      0.024 &     10.916 &      0.023 &          g &          1 \\
BPS CS 22894-0019 &23 39 19&+00 03 42&13.92 &     12.839 &      0.026 &     12.523 &      0.025 &     12.505 &      0.030 &          d &          1 \\
 HD 222434 &   23 40 42 &  -34 41 47 & 8.81 &      6.821 &      0.032 &      6.242 &      0.055 &      6.095 &      0.016 &          g &          1 \\
BD +18 5215 &  23 46 56 &  +19 28 22 & 9.74 &      8.795 &      0.032 &      8.560 &      0.031 &      8.535 &      0.018 &          d &          1 \\
BPS CS 29499-0060&23 53 40&-26 58 44 &13.05 &     12.102 &      0.024 &     11.855 &      0.024 &     11.865 &      0.021 &          d &          1 \\
\hline
\end{tabular} 
} 
\end{table*}

\end{document}